%% file: main.tex
\documentclass[10pt,twocolumn,letterpaper]{article}

\usepackage[accsupp]{axessibility} 
\usepackage[pagenumbers]{cvpr} 

\usepackage{times}
\usepackage{epsfig}
\usepackage{graphicx}
\usepackage{amsmath}
\usepackage{amssymb}
\usepackage{mathtools}
\usepackage{bm}
\usepackage{booktabs}
\usepackage{lipsum}
\usepackage{multirow}
\usepackage{tabularx}
\usepackage{makecell}
\usepackage{blindtext}
\usepackage{url}
\usepackage{enumitem}
\usepackage[normalem]{ulem}


\definecolor{cvprblue}{rgb}{0.21,0.49,0.74}
\usepackage[pagebackref=true,breaklinks=true,letterpaper=true,colorlinks,bookmarks=false]{hyperref}

\newcommand{\bs}{\boldsymbol}

\def\paperID{6143} 
\def\confName{CVPR}
\def\confYear{2025}

\begin{document}

\title{Mani-GS: Gaussian Splatting Manipulation with Triangular Mesh}

\author{
Xiangjun Gao$^1$ 
\quad 
Xiaoyu Li$^{2}$ $^{\dagger}$
\quad 
Yiyu Zhuang$^3$
\quad 
Qi Zhang$^2$ 
\quad 
Wenbo Hu$^2$
\quad 
Chaopeng Zhang$^2$  \\
\quad 
Yao Yao$^3$ $^{\dagger}$
\quad 
Ying Shan$^2$
\quad 
Long Quan$^1$\\
\\
$^1$The Hong Kong University of Science and Technology
\quad 
$^2$Tencent
\quad 
$^3$Nanjing University \\
\\
}

\twocolumn[\maketitle\vspace{-3em}\input{sec/teaser}\bigbreak]

\input{sec/0_abstract}    
\input{sec/1_intro}
\input{sec/2_related_work}

\input{sec/3_methods}
\input{sec/4_experiments}

\input{sec/5_conclusion}

\newpage
{
    \small
    \bibliographystyle{ieeenat_fullname}
    \bibliography{main}
}


\input{sec/X_suppl}

\end{document}

%% file: sec/teaser.tex
\begin{center}
\includegraphics[width=1\linewidth]{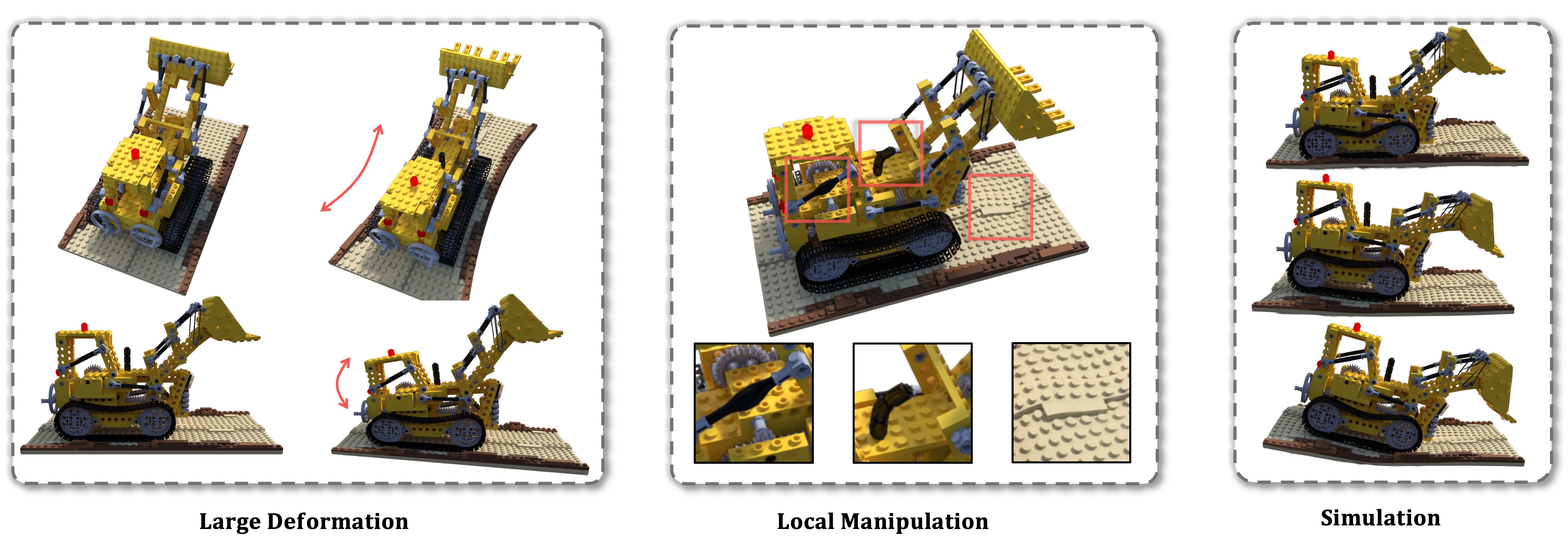}
\end{center} \vspace{-1.5em}
\captionof{figure}{Our proposed approach allows for 3DGS manipulation, including {\textit{large deformation}}, \textit{local manipulation}, and even \textit{physical simulation} (such as soft body), while maintaining high-quality rendering.
(Check \href{https://gaoxiangjun.github.io/mani_gs/}{\textbf{\emph{project page}}} for more visual results.)}
\label{fig:teaser}



%% file: sec/0_abstract.tex
\begin{abstract}




Neural 3D representations, such as Neural Radiation Fields (NeRF), excel at producing photorealistic rendering results but lack the flexibility for manipulation and editing which is crucial for content creation. 
%
%
However, manipulating NeRF is not highly controllable and requires a long training and inference time. 
With the emergence of 3D Gaussian Splatting (3DGS), extremely high-fidelity novel view synthesis can be achieved using an explicit point-based 3D representation with much faster training and rendering speed. However, there is still a lack of effective means to manipulate 3DGS freely while maintaining rendering quality. 
In this work, we aim to tackle the challenge of achieving manipulable photo-realistic rendering. We propose to utilize a triangular mesh to manipulate 3DGS directly with self-adaptation. This approach reduces the need to design various algorithms for different types of 3DGS manipulation. 
\let\thefootnote\relax\footnotetext{$^{\dagger}$Corresponding author: Xiaoyu Li and Yao Yao.}
By utilizing a triangle shape-aware Gaussian binding and adapting method, we can achieve 3DGS manipulation and preserve high-fidelity rendering. 
%
%
In addition, our method is also effective with inaccurate meshes extracted from 3DGS. 
Experiments demonstrate our method's effectiveness and superiority over baseline approaches.  
%


\end{abstract}

%% file: sec/1_intro.tex
\vspace{-15pt}
\section{Introduction}
\label{sec:intro}
Editing 3D content is essential for content creation and has various applications in movies, gaming, and virtual/augmented reality. 3D model editing enables users to create and modify models flexibly, thereby enhancing production efficiency. The traditional pipeline for modeling and editing a 3D asset with photo-realistic rendering involves a process with geometry modeling, texturing, UV mapping, lighting, and rendering, which is a tedious and time-consuming flow requiring lots of manual work.


Over the past few years, the neural radiance field (NeRF)~\citep{mildenhall2021nerf} has been widely studied due to its high capability and simple reconstruction process in 3D representation. However, the implicit representation poses challenges for editing. To address this, some methods are proposed to edit this implicit neural radiance field~\citep{mildenhall2021nerf}. NeRF-Editing~\citep{yuan2022nerf} is the first to utilize the triangular mesh to help edit the implicit radiance field. 
%
%
Volume rendering is conducted in the deformed space to render the deformed object by editing the triangular mesh. The sampling points in the deformed space are mapped to canonical space based on the constructed tetrahedra grid.
~\citep{jambon2023nerfshop, liu2023neural} present similar ideas by employing tetrahedra to deform sampling points and achieve editable rendering. In addition, NeuMesh~\citep{yang2022neumesh} and SERF~\citep{zhou2023serf} define the neural radiance field by associating each mesh vertex with radiance and geometry features. Then they conduct volume rendering like Point-NeRF~\citep{xu2022point} in deformed space without backward mapping. However, these editing methods based on implicit neural radiance fields still suffer from inconvenient manipulation, suboptimal rendering results, and long training and rendering times.

Recently, 3DGS~\citep{kerbl20233d} has gained significant attention in differential rendering due to its high-fidelity and fast rendering proficiency. However, despite being an explicit 3D representation, it still lacks an effective method for manipulating 3DGS while maintaining high-quality rendering. SuGaR~\citep{guedon2023sugar}, the work most closely related to our objective, develops a novel algorithm to extract a triangular mesh from 3DGS. Although their main goal is not to facilitate editable photorealistic rendering, they bind the 3DGS to the extracted mesh, enabling model animation.

This paper proposes a method that enables 3DGS manipulation, achieving high-quality and photo-realistic rendering. Our key insight is to manipulate 3DGS using a triangular mesh as the proxy, which allows for direct transfer of mesh manipulation to 3DGS with 3DGS self-adaptation. With our methods, we can achieve \textit{large deformation}, \textit{local manipulation}, and \textit{soft body simulaton} with high-quality results as shown in Fig \ref{fig:teaser}, which also avoid the need to design various algorithms for different types of manipulation. 

To achieve controllable 3DGS manipulation through mesh, an intuitive approach is to bind the GS to lay perfectly on the triangle and enforce the GS to be thin enough. After mesh manipulation, the GS will automatically adapt its rotation and position with the attached triangle, as employed in SuGaR~\citep{guedon2023sugar}. However, SuGaR heavily relies on the accuracy of mesh geometry, inheriting the defects of mesh rendering. Specifically, for inaccurate parts, SuGaR cannot inpaint the missing parts or remove the redundant parts during rendering. 

Adding an offset to the position of attached Gaussians during the reconstruction may seem like a reasonable solution to compensate for mesh inaccuracy. However, this fixed offset cannot be generalized well to the deformed space after manipulation. 
Our proposed solution is to define a local coordinate system for each triangle, which we refer to as \textit{local triangle space}. We then bind Gaussians to each triangle and optimize the Gaussian attributes, including rotation, position, and scaling, in the attached \textit{local triangle space}. 

During mesh manipulation, the attributes in the local triangle space remain unchanged, while the global Gaussian position, scaling, and rotation will be self-adaptively adjusted according to our proposed formula. As a result, our proposed approach enables us to manipulate 3DGS using a triangular mesh while maintaining rendering quality. Since our Gaussians are set to be free outside the triangle, we can also support high-fidelity manipulation even when the Gaussians are bound to an inaccurate mesh, exhibiting a high tolerance for mesh accuracy.

The contributions of our paper are listed as follows:
\begin{itemize}
    \item We propose a 3DGS manipulation method that can effectively transfer the triangular mesh manipulation to 3DGS and maintain high-quality rendering.
    \item We introduce a triangle shape aware Gaussian binding strategy with self-adaption, which has a high tolerance for mesh accuracy.
    \item We evaluate our method and achieve state of the art results, also supporting various 3DGS manipulations including \textit{large deformation}, \textit{local manipulation}, and \textit{soft body simulation}.
\end{itemize}

%% file: sec/2_related_work.tex
\section{Related Work}
\label{sec:relatedwork}
\subsection{NeRF Editing}
Recently, NeRF~\citep{mildenhall2021nerf} has garnered significant attention due to its high-quality and photo-realistic rendering results for novel view synthesis. NeRF represents the scene as a continuous function that maps a spatial location and viewing direction to a volume density and color, which is parameterized by a multilayer perceptron (MLP). Owing to the implicit representation that encodes the scene within the network parameters, editing and deforming the geometry of the NeRF scene explicitly like mesh can be challenging. To enable user editing of NeRF, 
~\citep{liu2021editing} introduce editing conditional radiance fields trained on a shape category. However, it only supports basic editing, such as removing/adding object parts or shape transfer. CLIP-NeRF~\citep{wang2022clip} achieves NeRF editing with text or images by leveraging CLIP model~\citep{radford2021learning} but still can not edit the geometry locally. Some other work~\citep{xiang2021neutex, wang2023seal3d, bao2023sine, zhan2023general} edit the NeRF in texture level which is not the focus of this paper. 

To edit and deform NeRF locally, ~\citep{jambon2023nerfshop,yuan2022nerf, liu2023neural} construct a tetrahedra grid based on the underlying 3D shape. After explicitly deforming the tetrahedra into the posed space for editing, the sampled 3D points are mapped from the posed space to the canonical space through the unaltered tetrahedron, which means the canonical position can be calculated from the shared barycentric coordinate for both deformed and canonical tetrahedron. The density and radiance in the posed space can be calculated for the mapped sampling points in canonical space. On the other hand, ~\citep{wang2023mesh, zhou2023serf, yang2022neumesh} employ mesh as the guidance for deformation. NeuMesh~\citep{yang2022neumesh} presents a novel representation to encode neural implicit field on a mesh-based scaffold for geometry and texture editing.~\citep{wang2023mesh} achieves the manipulation of both the geometry and color of neural implicit fields through differentiable colored meshes. Furthermore, there are several methods~\citep{sun2022fenerf, ma2022neural, zhang2022fdnerf, chen2023uv, zheng2022structured, zhuang2022mofanerf, wu2023high, zhuang2024towards, gao2022mps} that particularly focus on editing NeRF for avatar. In this work, we introduce an editing method based on 3DGS for general objects.


\subsection{Mesh-based NeRF Rendering}
NeRFs have shown impressive rendering results, however, rendering one pixel using NeRF necessitates a volumetric rendering algorithm that involves inferring MLP hundreds of times to estimate their radiance and density. 
To accelerate NeRF rendering, several methods~\citep{ rakotosaona2023nerfmeshing, yariv2023bakedsdf, chen2023mobilenerf, yao2022neilf,tang2023delicate, gao2024contex, yu2024hifi, he2024magicman} have been proposed to combine NeRF representation with mesh reconstruction. By converting this implicit representation to an explicit mesh, these methods may also facilitate applications like editing. In particular, MobileNeRF~\citep{chen2023mobilenerf} represents NeRF as a collection of polygons with deep feature textures, which can be rendered using the classic polygon rasterization pipeline.
%
%
Additionally, NeRFMeshing~\citep{rakotosaona2023nerfmeshing} distills the reconstructed NeRF into a signed surface approximation network to extract 3D mesh and shows the simulation results for editing. 
However, the editing results of these methods heavily rely on the accuracy of the extracted mesh. Artifacts present in the mesh will directly influence the editing results. In contrast, our method demonstrates robustness to the reconstructed mesh and can still yield promising results even with the inaccurate mesh.

\subsection{Gaussian Splatting Editing, Simulation and Animation}
3D Gaussian Splatting ~\citep{kerbl20233d} presents an innovative 3D Gaussian scene representation, accompanied by a differentiable renderer that attains real-time rendering of radiance fields while maintaining high quality. Initially, 3D Gaussian Splatting focuses solely on static scenes, which has been extended to model dynamic scenes~\citep{wu20234d, huang2023sc, yang2023deformable,lin2023gaussian, zhou2023drivinggaussian}, human avatars~\citep{zielonka2023drivable, qian2023gaussianavatars, yuan2023gavatar,  hu2023gaussianavatar, xu2023gaussian, kirschstein2023diffusionavatars, abdal2024gaussianshellmap}, Gaussian Splatting simulation and animation~\citep{guedon2023sugar, xie2023physgaussian, jiang2024vr, feng2024gaussian}. Specifically, SuGaR~\citep{guedon2023sugar} proposes a method to extract meshes from 3D Gaussian Splatting with additional regularization, in which they bind 3DGS on extracted mesh and animated with Gaussian Splatting rendering. GSP~\citep{feng2024gaussian}  incorporates physically-based fluid dynamics in 3DGS and PhysGaussian~\citep{xie2023physgaussian} introduces a unified simulation-rendering pipeline that generates physics-based dynamics with photorealistic renderings. VR-GS~\citep{jiang2024vr} achieves interactive physics-based editing in Virtual Reality. In this work, we also introduce a 3DGS manipulation method that binds Gaussian Splatting to the mesh, achieving state-of-the-art results.

Several concurrent works have emerged in parallel with ours.
In particular, GaMeS~\citep{waczynska2024games} constrains the 3DGS should be a flat ellipse and lay on the surface tightly, which means the position of 3DGS is inside the triangle and rotation is the same with triangle rotation; Mesh-GS ~\citep{gao2024mesh} permits an offset along the normal direction but the rotation and scaling are fixed after manipulation. Our model allows 3DGS to move out of the triangle and maintain the relative position and rotation after manipulation. And we can achieve high-quality rendering without the need for accurate mesh. 
Following SuGaR, Guedon et al.~\cite{guedon2024gaussian} proposed to represent the scene using a multi-layer mesh, while our method requires only a single-layer mesh.

%% file: sec/3_methods.tex
\section{Method}

Recently, due to its exceptional high-fidelity rendering capabilities and fast rendering speed, 3DGS~\citep{kerbl20233d} has emerged as a popular 3D representation in the differential rendering field. However, despite being an explicit 3D representation, it still lacks a way to manipulate this 3D representation for editing while maintaining high-quality rendering after the manipulation. In this work, giving multi-view RGB images of an object as input, we introduce a method for object manipulation that can achieve photo-realistic editable rendering by employing 3DGS.

\begin{figure*}[htpb]
	\centering
	\includegraphics[width=1.0\linewidth]{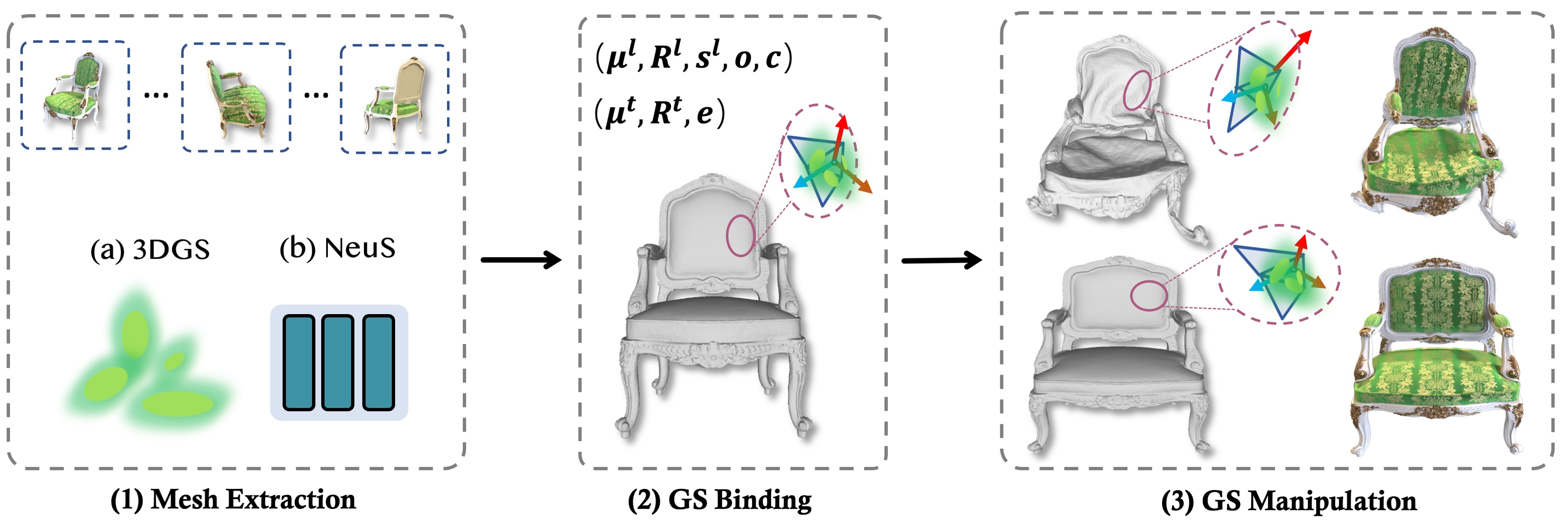}
        \caption{{{Overview of our method.}}(1) Firstly, we extract a triangular mesh from 3DGS~\citep{kerbl20233d} or a neural surface field(NeuS~\citep{wang2021neus}). (2) Next, we bind $N$ Gaussians to each triangle in the local triangle space, and optimize the local Gaussian attributes ($\bm {\mu^l, R^l, s^l, o, c}$). The triangle attributes ($\bm{\mu^t, R^t, e}$) are calculated based on the triangle vertices. (3) Finally, we manipulate 3DGS by transferring the mesh manipulation directly, thus achieving manipulable rendering.} 
        \vspace{-5pt}
        \label{fig:pipeline}
\end{figure*}

The pipeline of our method is illustrated in Figure~\ref{fig:pipeline} and consists of three main stages. First, we extract a mesh from 3D Gaussian Splatting (3DGS) or a neural surface field for subsequent 3D Gaussian binding (Sec.~\ref{sec:mesh}). Next, we devise a {novel} Mesh-Gaussian binding method dedicated to manipulating 3DGS while maintaining photo-realistic rendering quality (Sec.~\ref{sec:binding}). Finally, we describe the types of Gaussian manipulation we support(Sec.~\ref{sec:editing}). 

\subsection{Preliminary}

Thanks to its superior capability for high-fidelity and rapid rendering, 3D Gaussian Splatting (3DGS) ~\citep{kerbl20233d} has recently emerged as a popular 3D representation in differential rendering. 3DGS utilizes explicit 3D Gaussians as its primary rendering primitives. A 3D Gaussian point is mathematically defined as:

\vspace{-5pt}
\begin{equation}
G(\boldsymbol{x}) = exp(-\frac{1}{2}(\boldsymbol{x}-\boldsymbol{\mu})^\top\Sigma^{-1}(\boldsymbol{x}-\boldsymbol{\mu})) .
\label{eq:3d_gaussian}
\end{equation}

Each 3D Gaussian point is characterized by a 3D mean position coordinate $\boldsymbol{\mu}$ and a covariance matrix $\Sigma$. Additionally, each Gaussian has an opacity $\boldsymbol{o}$ and a view-dependent color $\boldsymbol{c}$  represented by a set of spherical harmonics (SH). To ensure that the covariance matrix $\Sigma$ retains its meaningful interpretation, it is parameterized as a unit quaternion $\boldsymbol{q}$ and a 3D scaling vector $\boldsymbol{s}$, defined as ${\Sigma = \boldsymbol{RS}\boldsymbol{S}^\top \boldsymbol{R}^\top}$.

To render an image from a specific viewpoint, 3D Gaussians are projected onto the image plane, resulting in 2D Gaussians. The 2D covariance matrix is approximated as: 

\begin{equation}
\Sigma'=\boldsymbol{J}\boldsymbol{W}\Sigma\boldsymbol{W}^\top\boldsymbol{J}^\top,
\end{equation}

where $\boldsymbol{W}$ and $\boldsymbol{J}$ denote the viewing transformation and the Jacobian of the affine approximation of perspective projection transformation~\citep{zwicker2002ewa}, respectively. The 2D means are calculated through the projection matrix. After this, the pixel color is composited through the alpha blending of $N$ ordered 2D Gaussians:
\vspace{-5pt}
\begin{equation}
\mathcal{C}=\sum_{i\in{N}}T_{i}\alpha_{i}\boldsymbol{c}_{i} \text{ with } \hspace{0.5em} T_i = \prod_{j=1}^{i-1}(1-\alpha_{i}) .
    \label{eq:alpha_blending}
\end{equation}
Here, $\alpha$ is obtained by multiplying the opacity $\boldsymbol{o}$ with the 2D covariance's probability computed from $\Sigma'$ and pixel coordinate on the image space. 


\subsection{Mesh Extraction}
\label{sec:mesh}
Our method can achieve high-quality editing using guided meshes obtained from various methods. In this section, we investigate different mesh extraction and reconstruction techniques with different mesh accuracy and processing time to guide the 3D Gaussians in our approach.


 \textbf{Marching Cube for Gaussian Splatting.} 
 In DreamGaussian~\citep{tang2023dreamgaussian}, the method attempts to summarize the alpha values of neighboring Gaussian points as the composite density value of marching cube sampling points. The mesh extracted using this method often ignores the thin and small structures.  With our Mesh-Gaussian binding strategy, we can achieve high-fidelity rendering with the inaccurate mesh and support smooth Gaussian manipulation.

  \textbf{Screened Poisson Reconstruction.} 3D Gaussian Splatting could be considered a type of point cloud, making it intuitive to extract the mesh using the Poisson-reconstruction algorithm. However, the 3D Gaussians do not have normal vectors for reconstruction. Inspired by recent 3DGS inverse rendering methods~\citep{gao2023relightable, liang2023gs}, we allocate an additional gaussian attribute, normal $\boldsymbol{n}$, for 3D Gaussians, which is supervised by the pseudo normal derived from depth map. After training 3DGS with normal attributes, we can extract the mesh using the Screened poisson surface reconstruction~\citep{kazhdan2013screened} algorithm. 


\textbf{Neural Implicit Surfaces.}
In this work, we also try to extract high-quality surfaces from the implicit representation utilizing the method proposed in NeuS~\citep{wang2021neus}. NeuS mesh has a large number of triangles, which negatively affects both training and inference speeds. We utilize mesh decimation techniques to reduce the count of triangles to approximately 300K.

\subsection{Binding Gaussian Splatting on Mesh}
\label{sec:binding}

Owing to the exceptional proficiency in high-fidelity and fast rendering, 3DGS has gained significant attention in differential rendering. However, despite being an explicit 3D representation, it currently lacks a method for effectively manipulating 3DGS while preserving high-quality rendering simultaneously. Mesh editing techniques, such as large-scale deformation, localized manipulation, and simulation, have been widely acknowledged and extensively researched for many years. Our primary objective is to associate the 3DGS with mesh triangles, enabling the manipulation of 3DGS and its rendering results following mesh editing.

Given a reconstructed or extracted triangular mesh $\boldsymbol{T}$ with $K$ vertices $\{\boldsymbol{v}_i\}_{i=1}^{K}$ and $M$ triangles $\{\boldsymbol{f}_i\}_{i=1}^{M}$, the goal of our method is to construct a 3DGS model bound to mesh triangles and optimize each Gaussian attribute $\{\boldsymbol{\mu}_i, \boldsymbol{q}_i, \boldsymbol{s}_i, o_i, \boldsymbol{c}_i\}$. To simplify the notation, we will omit the subscript in subsequent sections. 

For each triangle $\boldsymbol{f}$ in the given mesh $\boldsymbol{T}$, which is composed of three vertices $(\bm{v_1}, \bm{v_2}, \bm{v_3})$, we initialize $N=3$ Gaussians on this triangle. To be specific,
the mean position $\boldsymbol{\mu}$ of initialized Gaussians is formulated as $\boldsymbol{\mu}=(w_1\boldsymbol{v_1} +w_2\boldsymbol{v_2} +w_3\boldsymbol{v_3})$, $\bs{w} = (w_1, w_2, w_3)$ is the pre-defined barycentric coordinate of each Gaussians attached on the triangle. And $\bs{w}$ satisfy $(w_1 + w_2 + w_3) = 1$.

\vspace{3pt}
\textbf{Gaussians on Mesh.} To achieve controllable 3DGS manipulation through the mesh, an intuitive way is to perfectly attach 3DGS to the triangle, as shown in the SuGaR~\citep{guedon2023sugar}. With the rotation matrix denoted as $\bs{R} = \{\bs{r_1}, \bs{r_2}, \bs{r_3}\}$ and the scaling vector represented by $\bs{s} = (s_1, s_2, s_3)$, SuGaR train 3DGS with a flat Gaussian distribution on the mesh by setting $s_1 = \epsilon $, where $ \epsilon $ is close to zero. $\bs{r_1}$ is defined by the normal vector $\bs{n}$ of the attached triangle. The Gaussians have only 2 learnable scaling factors $(s_2, s_3)$ instead of 3, and only 1 learnable 2D rotation rather than a quaternion, to keep the Gaussians flat and aligned with the mesh triangles. 
This binding strategy is heavily dependent on mesh accuracy, which limits the flexibility of 3DGS in modeling complex object rendering. When the mesh accuracy is low, it cannot compensate for missing or redundant parts of geometry. Conversely, when the mesh accuracy is high but still deviates from the ground truth, it tends to produce a blurred effect due to the view inconsistency between the inaccurate mesh and multiview images. Overall, this binding strategy inherits the shortcomings of mesh rendering.
%

\vspace{3pt}
\textbf{Gaussians on Mesh with Offset.} To compensate for the inaccuracy of the extracted mesh, it would be better to add an offset $\Delta \bs{\mu}$ to the Gaussians 3D mean $\bs{\mu}$, which enables the Gaussians to move out of the attached triangle $\bs{f}$. Although it can improve the rendering quality of the reconstructed static object, this offset field is not generalized. It would result in noisy and unexpected rendering distortion in the manipulated object due to the mismatched relative position between 3DGS.

\vspace{3pt}
\textbf{Triangle Shape Aware Gaussian Binding and Adaption.} To preserve the high-fidelity rendering results after manipulation, the key lies in maintaining the local rigidity and preserving the relative location between Gaussians, both for 3D means and rotations. Our key insight is to define a local coordinate system in each triangle space. 

The first axis direction of triangle space is defined as the direction of the first edge.
The second axis direction of triangle space is defined as the triangle's normal direction.
The third axis direction of triangle space is defined as the cross product of the first and second axis.
Then the triangle coordinate system rotation can be formulated as:
\vspace{-5pt}
\begin{equation}
    \bs{R^t} = [\bs{r_1^t}, \bs{r_2^t}, \bs{r_3^t}] = [\frac{(\bs{v_2} - \bs{v_1})}{ \lVert \bs{v_2} - \bs{v_1} \rVert}, \bs{n^t}, \frac{(\bs{v_2} - \bs{v_1})}{ \lVert \bs{v_2} - \bs{v_1} \rVert} \times \bs{n^t}]
\end{equation}
\vspace{-5pt}

where $\bs{v_1}$, $\bs{v_2}$ is the first and second vertex location repectively, $\bs{n^t}$ is the normal vector calulated by:

\vspace{-5pt}
\begin{equation}
    \bs{n^t} =  \frac{(\bs{v_2} - \bs{v_1}) \times  (\bs{v_3} - \bs{v_1}) }{ \lVert (\bs{v_2} - \bs{v_1}) \times  (\bs{v_3} - \bs{v_1}) \rVert}. 
\end{equation}
\vspace{-5pt}

We then optimize the Gaussians' local position $\bs{\mu^{l}}$ and local rotation $\bs{R^l}$ in triangle space instead of the global position and rotation in the original 3DGS.

Then the global rotation, scale and location of 3DGS are as follows:

\vspace{-5pt}
\begin{equation}
    \bs{R = {R^t R^l}}, \bs{s} = {\bm s^l}, \quad \bs{\mu = R^t\mu^l + \mu^t}
\end{equation}
\vspace{-5pt}

where, $\bs{\mu^t}$ is the global coordinate of each triangle center. In practice, we initialize $N$ local Gaussian points and bind them for each Gaussian point, whose initialized position is on the triangle uniformly.

\begin{figure*}[th!]
	\centering
	\includegraphics[width=0.95\linewidth]{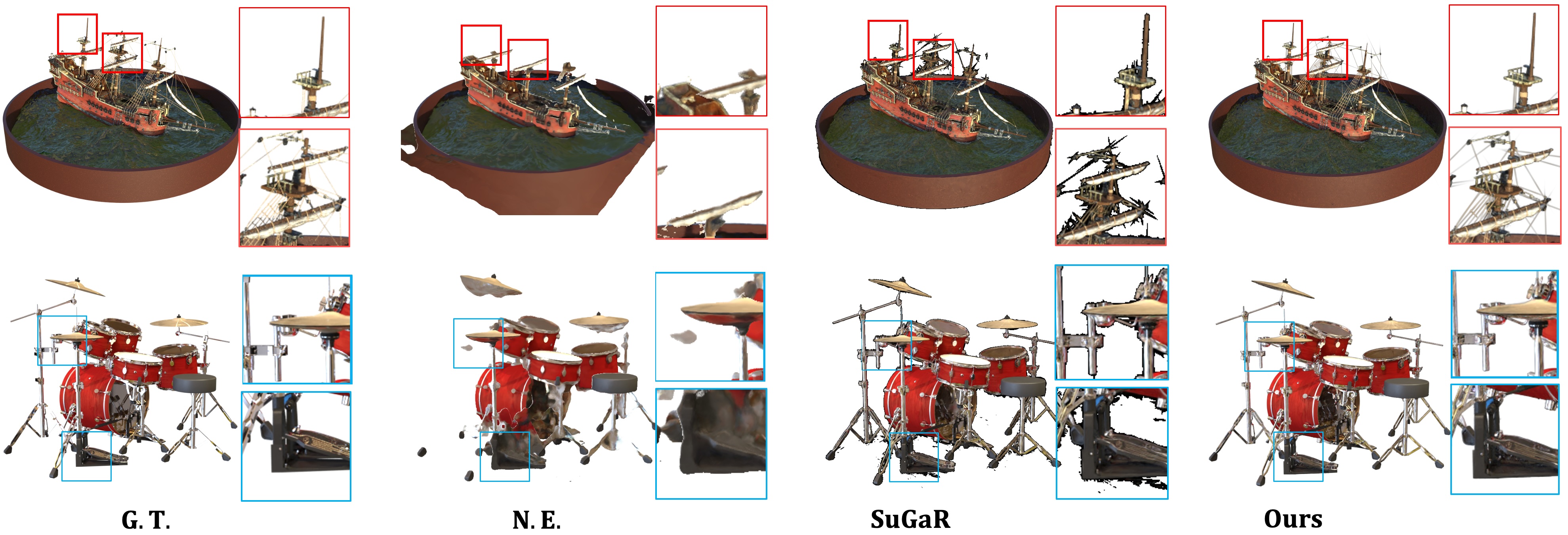}
	\vspace{-6pt}
        \caption{Visual comparison between ours, NeRF-Editing~\citep{liu2021editing}({N.E.}) and SuGaR~\citep{guedon2023sugar} for static rendering. It illustrates our proposed method can contain a much more accurate boundary in ``\textit{Ship}'', and distinct results in ``\textit{Drums}''.}
        \label{exp:comp_static}
\end{figure*}

\begin{table*}[htbp]

\setlength{\tabcolsep}{5pt}
\setlength{\aboverulesep}{0pt}
\setlength{\belowrulesep}{0pt}
\centering
\small

\caption{{
Quantitative comparison of our methods with  NeRF-Editing (``{\textit{N.E.}''})~\citep{yuan2022nerf} and {``\textit{SuGaR}''}~\citep{guedon2023sugar}} on NeRF Synthetic dataset in terms of SSIM, PSNR, LPIPS. 
``\textit{Ours}'' is with \textit{NeuS} mesh and ``\textit{Ours(S*)}'' is with the same mesh as ``\textit{SuGaR}''. 
The bests are marked in bold, second bests are underlined. ($\uparrow$ means higher is better, $\downarrow$ means lower is better)
}

\begin{tabular}{lcccc|cccc|cccc}
\toprule[1.pt]
\multicolumn{1}{c}{\multirow{2}{*}\textbf{Subject}} & \multicolumn{4}{c}{PSNR$\uparrow$}     & \multicolumn{4}{c}{SSIM$\uparrow$}     & \multicolumn{4}{c}{LPIPS$\downarrow$} \\
\cmidrule{2-13}      & \textit{N.E.} & \textit{SuGaR} & \textit{Ours(S*)} & \textit{Ours}  & \textit{N.E.} & \textit{SuGaR} & \textit{Ours(S*)} & \textit{Ours}  & \textit{N.E.} & \textit{SuGaR} &\textit{Ours(S*)} & \textit{Ours} \\
\midrule
Chair    &  28.15     & 34.23 &\textbf{35.68}    &\underline{35.38}&   0.943    & 0.984 & \textbf{0.987}  &\underline{0.986}&  0.061     & 0.013 &  \textbf{0.010} & \underline{0.011} \\
Drums    &\sout{21.14}& 25.78 &\underline{26.15} & \textbf{26.19}  &\sout{0.884}& 0.948 & \textbf{0.953}  &  \textbf{0.953} &\sout{0.120}& 0.044 &  \textbf{0.038} & \underline{0.039} \\
Ficus    &\sout{23.82}& 32.09 &\underline{35.29} & \textbf{35.40}  &\sout{0.909}& 0.973 & \textbf{0.986}  &  \textbf{0.986} &\sout{0.101}& 0.025 &  \textbf{0.013} & \textbf{0.013} \\
Hotdog   &   32.67    & 36.53 &\textbf{37.67}    &\underline{37.49}&   0.969    & 0.983 & \textbf{0.986}  &\underline{0.984}&  0.048     & 0.022 &  \textbf{0.017} & \underline{0.019} \\
Lego     &   29.16    & 35.05 &\textbf{36.41}    &\underline{36.33}&   0.944    & 0.979 & \textbf{0.983}  &\underline{0.982}&  0.074     & 0.018 &  \textbf{0.014} & \underline{0.015} \\
Material &  {29.48}   & 28.65 &\underline{29.51} & \textbf{29.91}  &   0.944    & 0.941 &\underline{0.951}&  \textbf{0.956} &  0.063     & 0.055 &  \textbf{0.042} & \underline{0.046} \\
Mic      &  {29.60}   & 35.39 &\underline{36.99} & \textbf{37.46}  &   0.952    & 0.990 & \textbf{0.992}  &  \textbf{0.992} &  0.046     & 0.011 &\underline{0.008}& \textbf{0.007} \\
Ship     &  {25.01}   & 28.55 &\underline{30.48} & \textbf{31.01}  &   0.083    & 0.869 &\underline{0.889}&  \textbf{0.890} &  0.194     & 0.124 &  \textbf{0.097} & \textbf{0.097} \\
\midrule
Average &  \sout{ }   & 32.06 &\underline{33.52} & \textbf{33.65} &\sout{    } & 0.958 &\textbf{0.966} & \textbf{0.966} &\sout{   }  & 0.039 &\textbf{0.030} & \underline{0.031} \\
\bottomrule[1.pt]
\end{tabular}%
\label{tab:comp}%
\end{table*}%

This binding strategy allows 3DGS to move outside the triangle while preserving the relative position and rotation of the Gaussians after mesh manipulation.

However, following mesh manipulation, not only does the triangle center change but also the triangle shape. With the altered triangle shape, the local Gaussian position and scaling should adjust accordingly. When the triangle enlarges, it is intuitive that the local scaling and position should expand as well: 

\vspace{-5pt}
\begin{equation}
    \bs{R = {R^t R^l}}, \bs{s} = {\beta \bm e \bm s^l}, \quad \bs{\mu = eR^t\mu^l + \mu^t}, 
    \label{eq:bind}
\end{equation}
\vspace{-5pt}

where $\beta$ is a hyper-parameter, \textit{adaption vector} $\bm e = [e_1, e_2, e_3]$ is designed to make sure that the global scaling $\bm s$ is proportionable to the triangle shape. The first axis is along the first edge, so $e_1$ is designed as the length $l_1$ of the first edge of the triangle. The second axis is along the normal direction, we set $e_2 = (0.5 * (e_1 + e_3))$. The third axis is perpendicular to the first edge, we set  $e_3$ as the average length of the second and third edges $(0.5 * (l_2 + l_3))$. 


\subsection{Manipulate Gaussian Splatting through Mesh}
\label{sec:editing}
Utilizing our triangle shape aware gaussian binding and adapting strategy, upon the completion of model training and mesh manipulation, the 3DGS is instantly manipulated and adapted. 
During mesh manipulation, the attributes in the local triangle space remain unchanged. The triangle rotation, position, and edge length can be calculated instantly. Therefore, the global Gaussian position, scaling, and rotation can be self-adaptively adjusted following our proposed formula.
In this paper, we exhibit the 3DGS manipulation rendering outcomes, such as large-scale deformation, local manipulation, and soft-body simulation, which are driven by the manipulated mesh. 
In our experiments, we employ Blender to manipulate the mesh.

%% file: sec/4_experiments.tex
\section{Experiments}

\subsection{Training Details}

\begin{figure*}[htpb]
	\centering
	\includegraphics[width=1.0\linewidth]{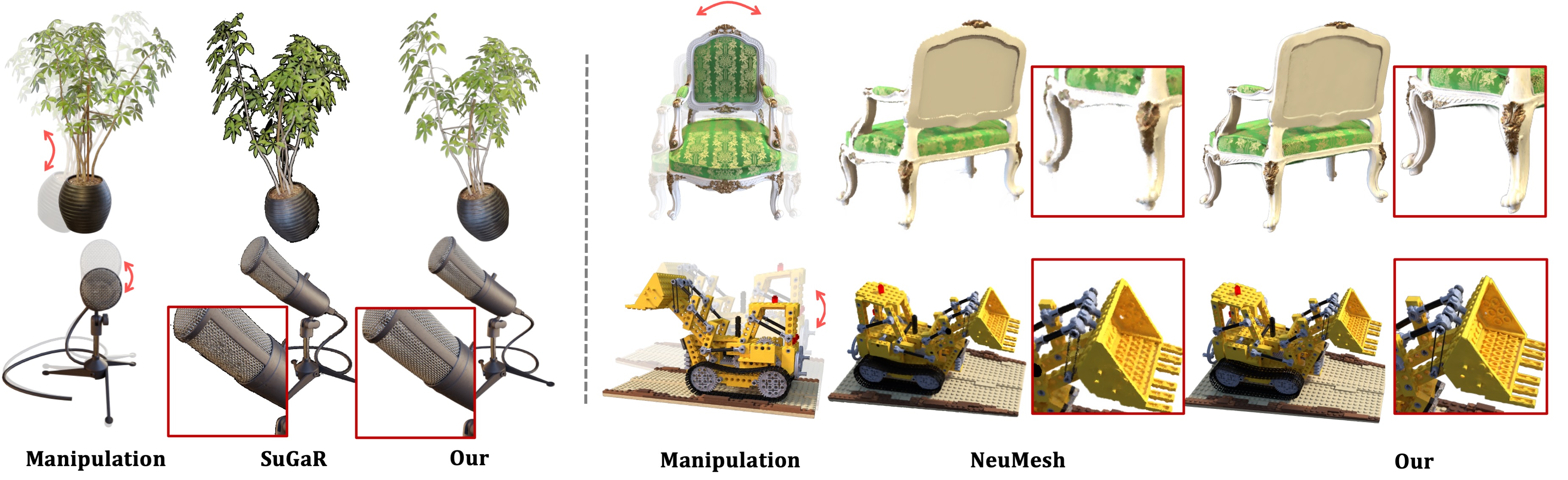}
	\vspace{-15pt}
        \caption{
        We offer an editing comparison between our method, SuGaR, and NeuMesh. Our approach demonstrates fewer artifacts and less blurring effects than SuGaR, and presents more abundant and distinct details compared to NeuMesh. For further details, please zoom in.
        }
        \label{exp:comp_edit}
	\vspace{-5pt}
\end{figure*}

The first stage of our methods includes a mesh extracting stage, during which we extract triangular mesh from NeuS~\citep{wang2021neus} or 3DGS (Screened Poisson or Marching Cube). However, the extracted mesh always contains enormous triangles, which we try to decimate to around 300K.

With the extracted mesh, we conduct the \textit{triangle shape aware Gaussian binding and adapting} strategy on the mesh. For each triangle, we bind $N=3$ Gaussian on the surface initially. The Gaussian attributes are optimized subsequently with the supervision of multi-view rendering loss in the second stage.
We train our model for 30K iterations in the initial stage to extract mesh and 20K iterations in the second stage. All experiments are conducted on a single NVIDIA A100 GPU. 


\vspace{-5pt}
\subsection{Datasets, Metrics and Methods for Comparison}
To evaluate our methods, we compare Mani-GS with previous editable novel view synthesis methods, NeRF-based editing method NeRF-Editing ~\citep{yuan2022nerf}, NeuMesh ~\cite{yang2022neumesh} and 3DGS-based editing method SuGaR~\citep{guedon2023sugar}. For the evaluation, we employ the commonly used metrics:\textit{PSNR}, \textit{SSIM}, \textit{LPIPS}. We evaluate our methods mainly on the NeRF Synthetic dataset~\citep{mildenhall2021nerf} and DTU dataset~\citep{jensen2014dtu}. 

\subsection{Evaluation}

\textbf{Static Rendering} Table \ref{tab:comp} provides a quantitative comparison of all NeRF Synthetic 8 cases between our method and competing methods. We conducted experiments using their official code repository. 
We remove two outliers (``\textit{Drums, Ficus}'') from NeRF-Editing using a strikeout. As observed, our approach outperforms all baseline methods in terms of PSNR, SSIM, and LPIPS, indicating that we achieve the best rendering quality. Our method achieves SOTA results across all 8 cases, surpassing SuGaR by at least 1.5 points in PSNR when using NeuS mesh. For a fair comparison with SuGaR, we also evaluate our binding method using the same mesh as SuGaR (denoted as ``\textit{Ours(S*)}''). Our method maintains its superior performance even with SuGaR mesh, showing comparable results to those with \textit{NeuS} mesh.

In Figure \ref{exp:comp_static}, we present qualitative results of our approach and other methods in overview and zoom-in details. For SuGaR, it attempts to bind 3D Gaussians on the triangle and enforce that the attached 3DGS is as flat as possible and is closely aligned with the triangle. 
This binding strategy heavily depends on the accuracy of the mesh. As can be observed in the third column of Figure \ref{exp:comp_static}, wrong geometry leads to an inaccurate rendering, especially in the boundary region. \textit{N.E.}\cite{liu2021editing} results are blurred and lack intrinsic details.  In contrast, our method achieves higher fidelity rendering with more precise boundaries and richer details.

\begin{table}[htbp]
\centering
\resizebox{1.0\linewidth}{!}{
\tabcolsep 13pt
\begin{tabular}{lccc}
\toprule
\multicolumn{1}{c}{\multirow{2}{*}{Methods}} & \multicolumn{3}{c}{DTU}  \\ 
\cmidrule(lr){2-4} 
\multicolumn{1}{c}{} & \multicolumn{1}{l}{PSNR $\uparrow$} & \multicolumn{1}{l}{SSIM $\uparrow$} & \multicolumn{1}{l}{LPIPS $\downarrow$} \\ \hline

NeuMesh~\citep{yang2022neumesh} & {28.29} & {0.921} & {0.117} \\
SuGaR~\citep{guedon2023sugar} & 30.06 & 0.921 & 0.136  \\
Ours & \textbf{31.50} & \textbf{0.943} & \textbf{0.088} \\
\bottomrule
\end{tabular}
}
\caption{
We compare quantitative rendering quality with SuGaR and NeuMesh on the DTU dataset. Note that ours and SuGaR employ the same underlying mesh extracted from \emph{poisson recon.}
}
\label{tab:dtu}
\end{table}

We also evaluate our methods on the DTU dataset \cite{jensen2014dtu} with real objects, as shown in Table \ref{tab:dtu}. It is important to note that we use the same base mesh as SuGaR in this dataset. As listed, we outperform both SuGaR and NeuMesh across all metrics, further demonstrating the efficacy of our 3DGS binding strategy.

\begin{figure}[h!]
         \centering
	\includegraphics[width=1.0 \linewidth]{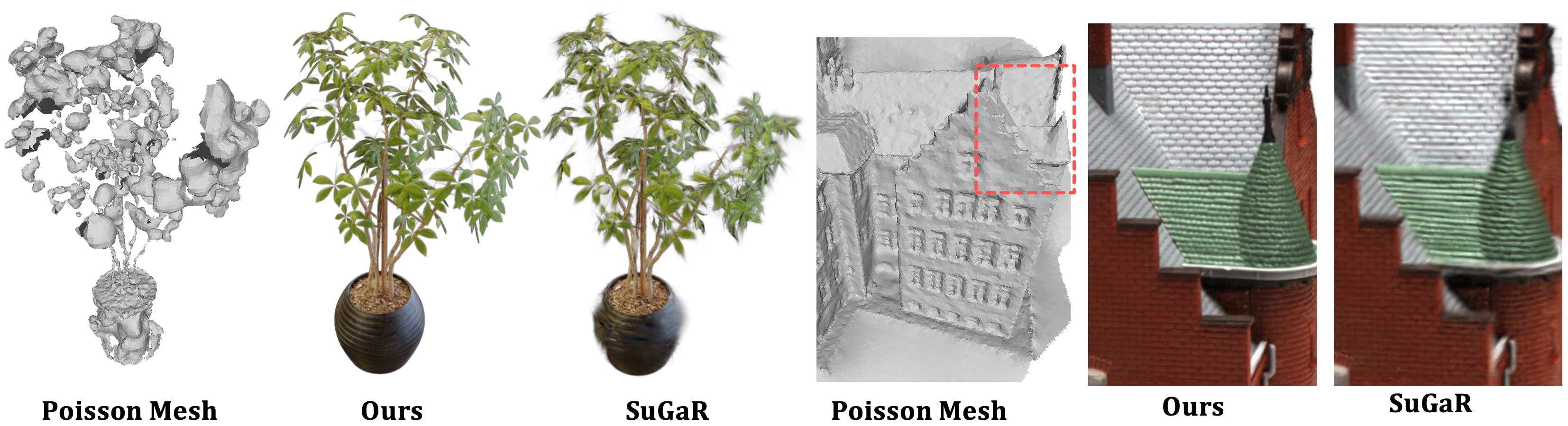}
	\vspace{-15pt}
        \caption{Visual comparison of inaccurate mesh binding and rendering. Note that ours and SuGaR employ the same underlying mesh extracted from \emph{poisson recon.} }
	\vspace{-5pt}
        \label{exp:sugar}
\end{figure}

\noindent\textbf{Binding 3DGS on inaccurate mesh.} 
When the mesh accuracy is low, as shown on the left side of Fig \ref{exp:sugar}, SuGaR produces distorted results due to the missing geometry. 
Conversely, when the mesh accuracy is high but still deviates from the ground truth mesh, SuGaR tends to generate blurred results because of view inconsistency with the given mesh.
Thanks to our binding strategy in the triangle's local space, which allows 3DGS to move freely around the triangle, we can achieve high-fidelity rendering in both scenarios. This demonstrates that our method exhibits a high tolerance for mesh accuracy.


\begin{figure}[htpb]
	\centering
	\includegraphics[width=1.0 \linewidth]{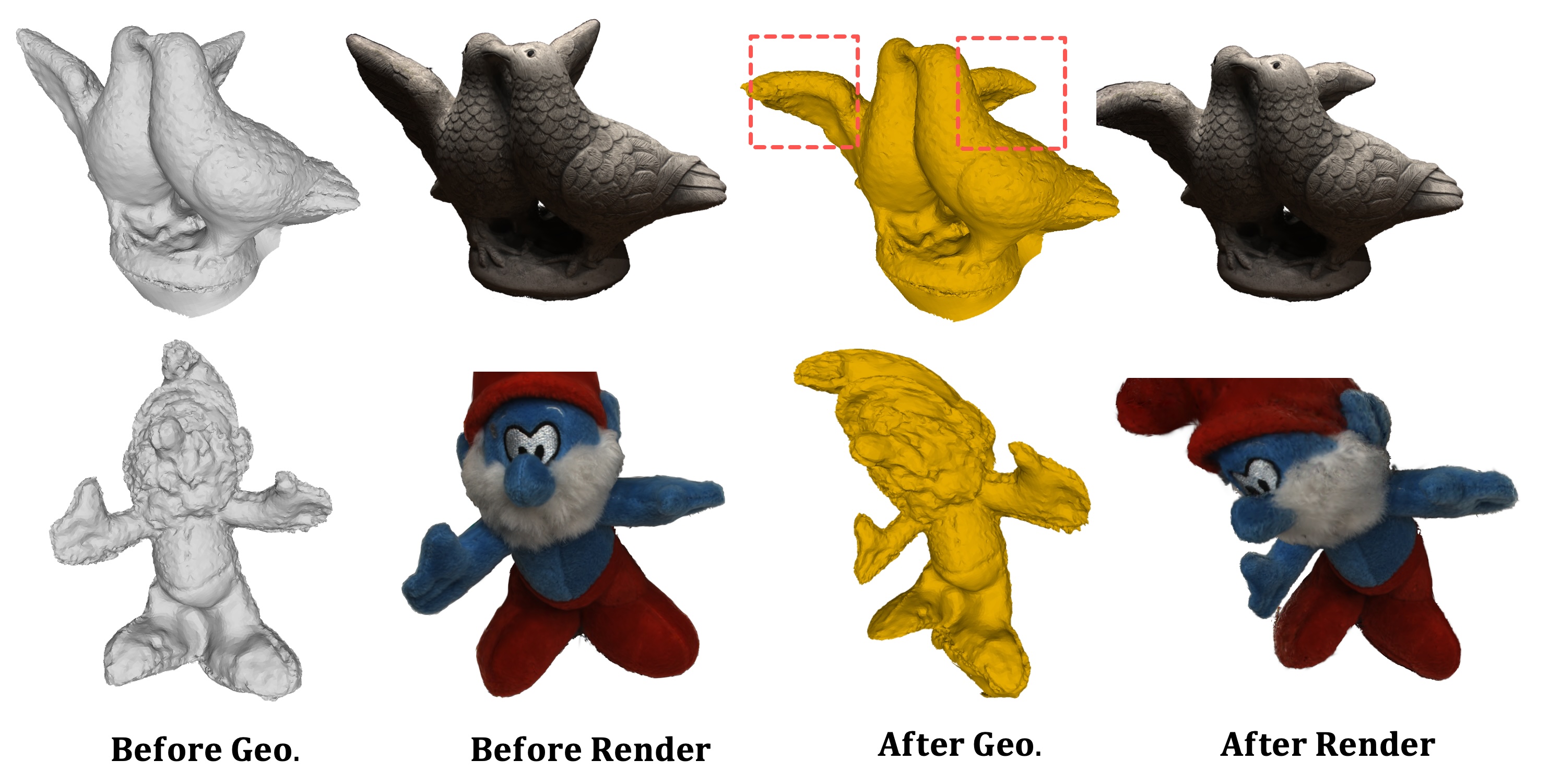}
        \caption{The manipulated geometry and corresponding rendering results of our methods on DTU dataset. }
	\vspace{-5pt}
        \label{exp:dtu}

\end{figure}

\noindent \textbf{Manipulation Rendering} In Figure \ref{exp:comp_edit}, we showcase our manipulation results. In these four cases, we manipulate the underlying mesh with large deformation, the \textit{Chair} is \emph{stretched}, \textit{Lego} is \emph{tapered}, \textit{Ficus} and \textit{Mic} is \emph{bent} respectively. As demonstrated in \textit{Ficus} and \textit{Mic}, we have more accurate boundaries and shapes, as well as distinct details. However, SuGaR can not adapt to compensate for geometry errors, which results in missing or dilated boundary rendering with distortions. For \textit{Lego} and \textit{Chair}, we can maintain the high rendering quality even after the large deformation, while NeuMesh showcases blurred and noised results.

\begin{figure}[htbp]
	\centering
	\includegraphics[width=1\linewidth]{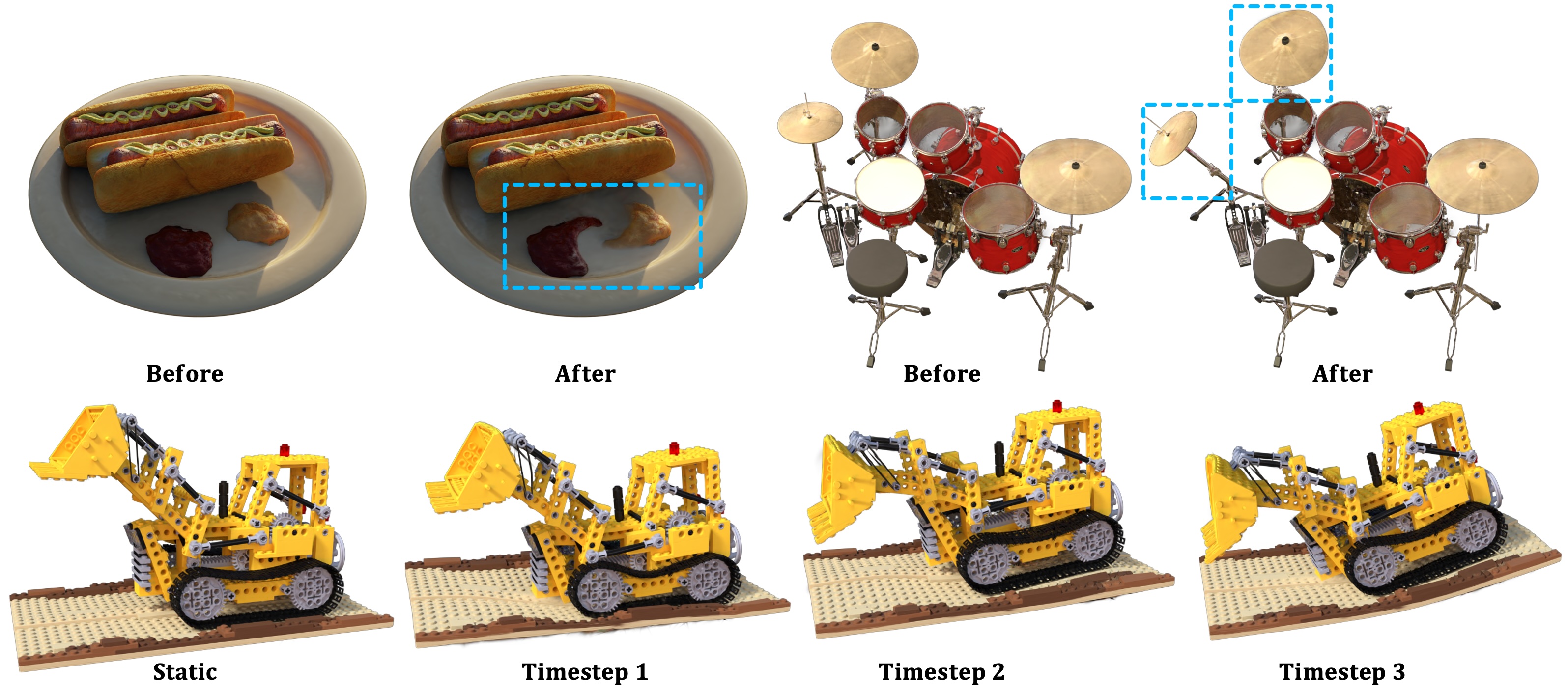}
	\vspace{-10pt}
        \caption{Visual results of {\textit{local manipulation} (row 1) and \textit{soft body simulation} at different timesteps. 
        }}
        \label{exp:comp_local}

\end{figure}

In addition to the large deformation, our method also produces promising results for local manipulation and physics simulation. Here we show an example of soft body simulation. In Figure \ref{exp:comp_local} row 1, we try to \textit{blend} the red sauce and yellow sauce of \textit{Hotdog} as shown in the blue box, which shows satisfying editing and reasonable rendering quality. In \textit{Drums}, we \textit{repose} a cymbal and \textit{elastically deform} a cymbal as shown in the blue box. After reposing and elastic deformation, we still preserve the photo-realistic rendering results. Note that the manipulation is achieved by manipulating the triangular mesh directly, the 3DGS rendering is achieved simultaneously with self-adaption.

\vspace{-5pt}

In row 2 of Figure \ref{exp:comp_local}, we present the rendering results of soft body simulation at different timesteps. As observed, we can achieve soft body simulation by just transferring the mesh simulation to 3DGS, which eliminates the need for a soft body simulation algorithm dedicated to 3DGS.

\subsection{Ablation Study}

\begin{figure}[htpb]
	\centering
	\includegraphics[width=1.0 \linewidth]{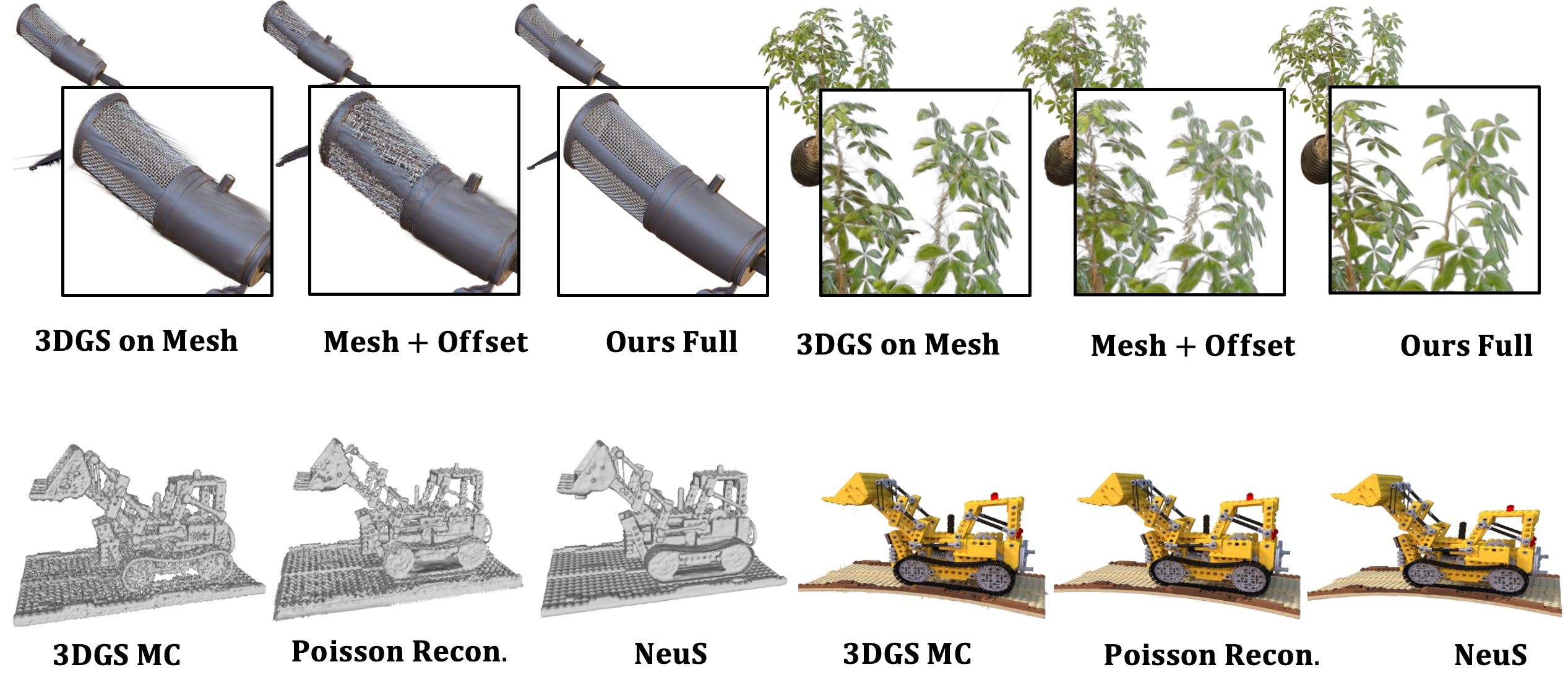}
        \caption{Ablation of binding strategy and geometry. After deformation, \textbf{{(3DGS on Mesh)}} shows a burring boundary, \textbf{(Mesh + Offset)} leads to significant noise and distortion, \textbf{(Ours Full)} can maintain the high fidelity rendering. In the second row, we demonstrate that even with a low-quality mesh, we can still achieve high-quality editable rendering.
        }
        \label{exp:abation}

\end{figure}

We conduct ablation studies to verify the effectiveness of triangle shape-aware Gaussian binding and adapting method. We first evaluate the strategy of directly binding \textbf{3DGS on mesh}, which implies that the 3D position is fixed on the triangle. The mesh is extracted from NeuS. As shown in Table \ref{tab:ab}, the performance significantly drops, with a decrease of approximately 2.6 PSNR compared to our best model. For visual ablation in Figure \ref{exp:abation}, \textbf{3DGS on Mesh} after deformation shows a boundary with many burrs. 

Next, we verify the effectiveness of adding 3D offset for 3DGS on Mesh (\textbf{Mesh + Offset}). Although the offset can enhance the fitting of 3DGS to the static scene, as demonstrated in Table \ref{tab:ab}, it fails to generate satisfactory deformation rendering results because the offset only fits the static scene and remains unchanged during subsequent deformations. Consequently, it leads to significant noise and distortion after manipulation in Figure \ref{exp:abation}, whereas our full model can maintain high-quality rendering after manipulation, demonstrating the efficacy of the 3DGS binding strategy in local triangle space.

\begin{table}[htbp]
\centering
\small
\caption{Quantitative ablation comparison between \textbf{3DGS On Mesh}, \textbf{Mesh + Offset}, \textit{Ours with Marching Cube Mesh or Screened Poisson Mesh} on NeRF Synthetic dataset. {($\uparrow$ means higher is better, $\downarrow$ means lower is better.)}} 
 \label{tab:ab}
\setlength{\tabcolsep}{0.8 mm}{
    \begin{tabular}{lccc}
        \toprule
        \multicolumn{1}{c}{\multirow{1}{*}{Method}} & PSNR$\uparrow$ & SSIM$\uparrow$ & LPIPS$\downarrow$ \\
        \midrule
3DGS On Mesh            & 30.87 & 0.9521 & 0.0447 \\ Mesh + Offset  & 32.48 & 0.9625 & 0.0341 \\
Ours + Marching Cube Mesh  &{32.11}&{0.9602}&{0.035} \\
Ours + Screened Poisson Mesh & {33.42} & {0.9638} & {0.0324} \\
Ours + NeuS Mesh & \textbf{33.45} & \textbf{0.9646} & \textbf{0.0309} \\
        \bottomrule
    \end{tabular}%
    }	
\end{table}

Finally, we conduct experiments with different meshes extracted using different methods. As can be observed in Figure \ref{exp:abation} row 2, 3DGS Marching Cube Mesh (\textbf{3DGS MC}) is of low quality, including a dilated boundary and very noisy surface. And the screened poisson mesh (\textbf{Poisson Recon.}) has some unconnected regions and missing parts compared with NeuS mesh. 
By employing our triangle shape aware Gaussian binding and adapting method, we can still achieve 3DGS manipulation and maintain high-fidelity rendering after large deformation. 
In Table \ref{tab:ab}, the numerical results obtained with screened poisson mesh are only slightly lower than those obtained with NeuS mesh. When the mesh is of low quality {(Marching Cube mesh)}, the quantitative results are approximately 1.3 PSNR lower than the best, but still 1.3 PSNR higher than only binding 3DGS inside the triangle of the best Mesh.


%% file: sec/5_conclusion.tex

\section{Conclusion}
\vspace{-5pt}
In this paper, we introduce a triangle shape aware Gaussian binding strategy with self-adaptation, which supports various 3DGS manipulations, maintains rendering quality, and has a high tolerance for mesh accuracy. We evaluate our methods on both synthetic and real datasets and demonstrate SOTA results. 

During our experiments, we noticed that some results still exhibit distortions. When the local region of the manipulated mesh contains highly non-rigid deformations, it can result in render distortions. Additionally, during our simulation demos, we found that conducting physics simulations on meshes with more than 35K triangles can take hours. 
%

%% file: sec/X_suppl.tex
\clearpage
\setcounter{page}{1}
\maketitlesupplementary

\textbf{Overview.} The supplementary material has the following contents:

\begin{itemize}[nosep,left=1.5em]
    \item More visual results
    \item Efficiency Analysis
    \item Implementation Details
\end{itemize}

\section*{A. More Visual Results}
\noindent\textbf{Demo Video.} In order to further demonstrate the effectiveness of our methods, we have provided additional visual videos showcasing large deformation, soft body simulation, and local manipulation. These videos can be accessed through \href{https://gaoxiangjun.github.io/contex\_human}{{\emph{project page}}}.

\begin{figure}[htpb]
	\centering
	\includegraphics[width=1.0\linewidth]{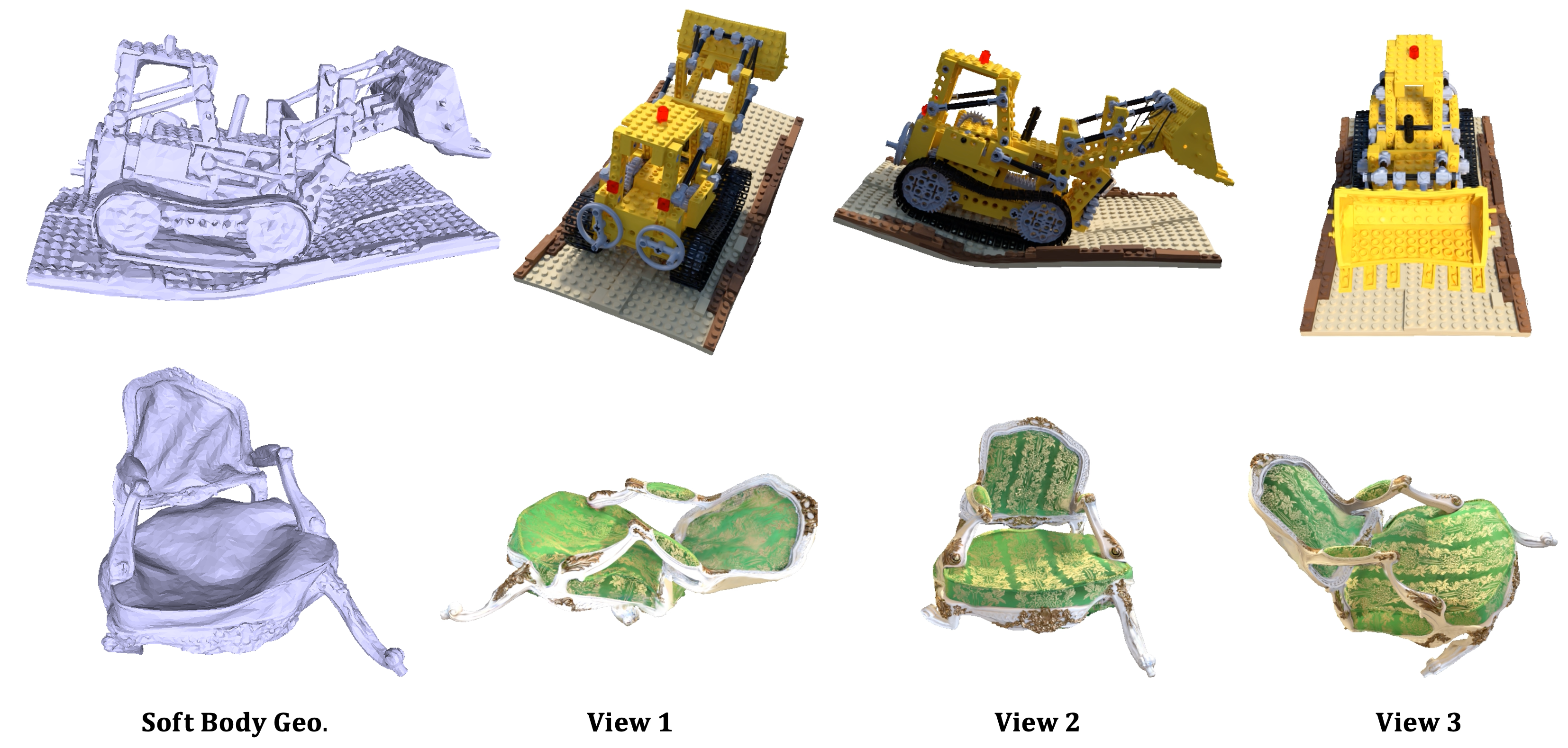}
        \caption{Visual results of softbody simulation at different viewpoints. The left column displays the geometry after simulation, while the right three columns showcase the rendering results from three different viewpoints.}
        \label{supple:soft_body}
\end{figure}

\label{sec:NeRF}
\noindent\textbf{Soft Body Simulation.} 
In addition to the visual results presented in the main paper, we also provide the geometry after simulation and rendering at different viewpoints in Figure \ref{supple:soft_body}. To improve the speed of the mesh simulation, we decimated the original mesh from 300K to 35K triangles. While this may result in some decrease in rendering quality due to the reduced number of triangles as well as Gaussians, it was necessary to ensure reasonable simulation speed.

\begin{figure}[htbp]
    \centering
    \includegraphics[width=1.0\linewidth, trim={0 0 0 0}, clip]{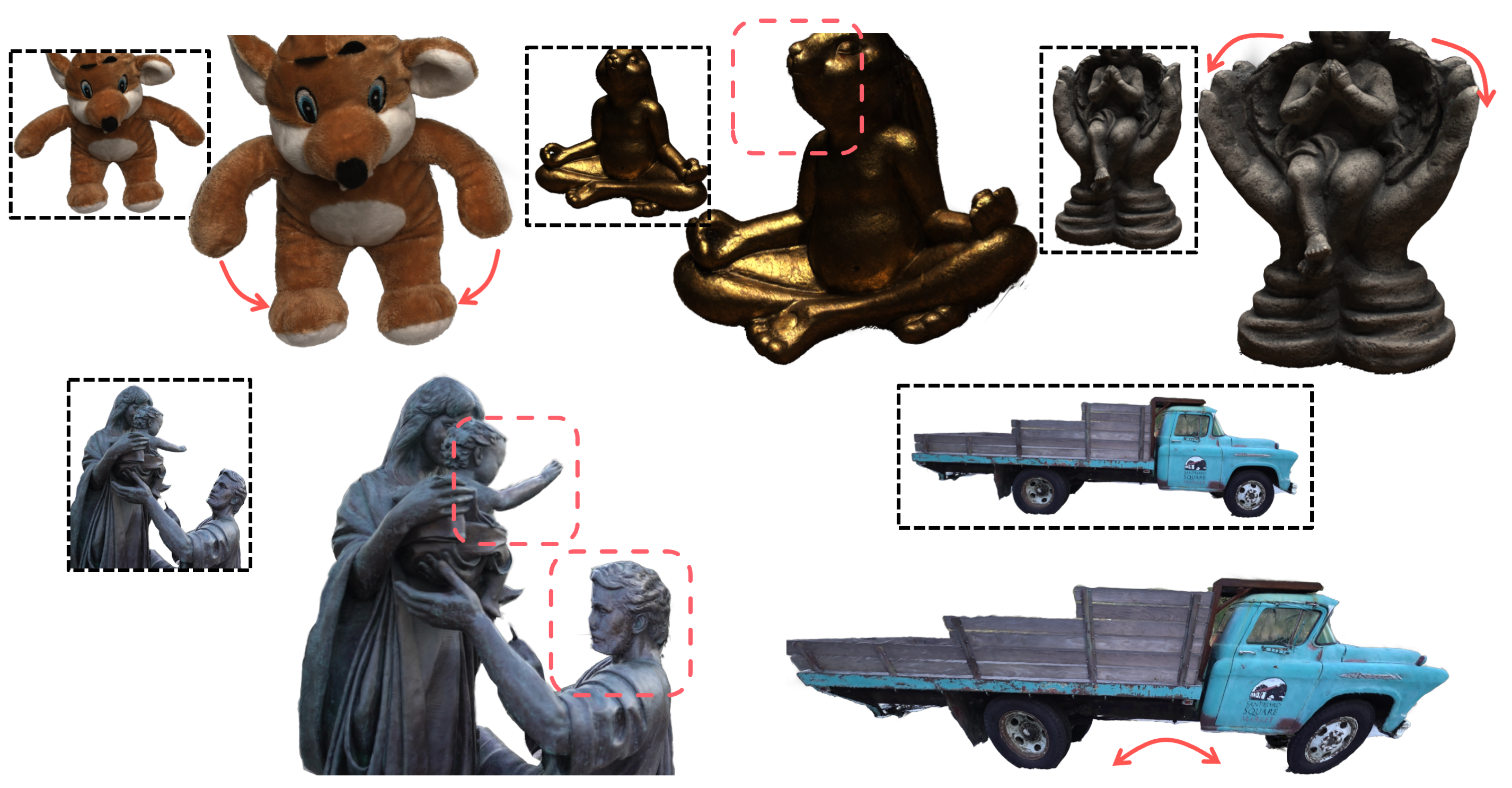}
    \caption{
    {Manipulation rendering results on real object dataset.  To highlight the deformed area, we have enclosed it within a red rectangle.}
    }
    \label{fig:dtu}
\end{figure}

\noindent\textbf{More real case.} We have also presented more manipulation results of real scenes in Figure \ref{fig:dtu}. 
The top three images are from the DTU dataset, while the bottom two are from the Tanks and Templates dataset\cite{Knapitsch2017_tnt}. Images within the dotted black rectangles are before manipulation, whereas those marked with red arrows or rectangles are after manipulation.
The manipulation results presented in Figure \ref{fig:dtu} demonstrate that our approach can successfully transfer mesh manipulation to Gaussian-Splatting, resulting in accurate and visually appealing results.

\begin{table}[htpb]
\centering
\small
\caption{Efficiency Analysis}
\label{efficiency}
\setlength{\tabcolsep}{0.8 mm}{
    \begin{tabular}{l|ccc|cc}
        \hline
    {} & $N$=4 & $N$=3 & $N$=1 & $N$=1 & $N$=1 \\
    \hline
    \textit{Triangles(K)} & 270 & 270 & 270 & 150 & 70 \\ 
    \textit{Points(K)} & 1080 & 810 & 270 & 150 & 70 \\ 
    \textit{Training (min)} & 16 & 13 & \textbf{7}  & 5.5 & 4.5 \\ 
    \textit{Speed (FPS)} & 244 & 300 & \textbf{452} & 571 & 572 \\ 
    \textit{PSNR} & 36.36  & \textbf{36.39} & 36.27 & 35.86 & 34.52 \\ 
    \hline
    \end{tabular}%
    }	
\end{table}

\section*{B. Efficiency Analysis}

\label{sec:effi}
The efficiency of 3DGS Binding training and rendering speed depends on the number of Gaussians, which is the product of the triangle number $T$ and the Gaussians number for each triangle $N$. In Table \ref{efficiency}, We first fixed $T$ and tested different values of $N$. Our results indicate that $N$=3 leads to the best rendering quality while keeping a competitive rendering speed. When $N=1$, the PSNR slightly decreased with a faster training and rendering speed.


We also evaluated the impact of underlying mesh resolution by testing meshes with different triangles (\textit{270K, 150K, 70K}). As shown in Table \ref{efficiency}, the rendering quality decreases while efficiency improves with decreasing mesh resolution. 

Regarding the editing time, it primarily depends on the time cost of mesh editing. We use Blender for mesh editing, and in our experience, \textit{local manipulation} and \textit{large deformation} can be achieved instantly. \textit{Soft body simulation} can be a more time-consuming process, as it depends on the simulation algorithm employed in Blender.

\section*{C. Implementation Details}
\label{sec:details}

\subsection*{C.1 Training Details of Mesh Extraction Stage}

As outlined in our main paper, the first stage of our approach involves mesh extraction. While we utilize the NeuS\cite{wang2021neus} mesh as the foundation for binding Gaussians, we also explore extracting mesh from Gaussian-Splatting.

In this work, we try to extract triangular mesh using the Screened Poisson surface reconstruction \cite{kazhdan2013screened} method from a trained Gaussian-Splatting model.  We incorporate a normal attribute $\bm n$ for each 3D Gaussian and optimize the normal attribute with the pseudo-normal constraint. 

The normal consistency is quantified as follows:
\begin{equation}
    \mathcal{L}_{n}=\|\mathcal{N}-\tilde{\mathcal{N}}\|_2 .
    \label{eq:cost_normal}
\end{equation}
where ${\mathcal{N}}$ is the rendered-normal map, $\tilde{\mathcal{N}}$ is the pseudo-normal map computed from rendered depth map.

Besides the normal constraint $\mathcal{L}_{n}$, the ordinary L1 Loss and Structural Similarity Index (SSIM) loss are also incorporated into optimization by comparing the rendered image $\mathcal{C}$ with the observed image $\mathcal{C}_{gt}$. To address the issue of unwarranted 3D Gaussians in the background region, we employ a mask cross-entropy loss. This loss is defined as follows:

\begin{equation}
\begin{split}
    \mathcal{L}_{mask} = -B^m\log{B} - (1-B^m)\log{(1-B)},
\end{split} 
\end{equation}
where $B^m$ denotes the object mask and $B$ denotes the accumulated transmittance $B=\sum_{i\in{N}}T_{i}\alpha_{i}$.

Then all the loss terms can be summarized as follows:
\begin{equation}
\begin{split}
    \mathcal{L}_{stage1} = \lambda_1\mathcal{L}_1 + \lambda_2\mathcal{L}_{SSIM} + \lambda_3\mathcal{L}_{n} + \lambda_4\mathcal{L}_{mask},
\end{split} 
\end{equation}
where $\lambda_1=1, \lambda_2=0.2 ,\lambda_3=0.01 , \lambda_4=0.1 $. We train this stage for 30K steps with adaptive density control, which is executed at every 500 iterations within the specified range from iteration 500 to 10K. Once the training stage is complete, we proceed with Screened Poisson surface reconstruction using the positions and normals of the Gaussians as input. The mesh extraction process takes less than 1 minute to complete.

In addition to mesh extraction, we also utilize Gaussian-Splatting Marching-Cube to extract the triangular mesh. Our approach involves sampling a grid with a resolution of $256 \times 256 \times 256$. For each sampling point, we identify its nearest Gaussian points. Sampling points that have the nearest Gaussians within a pre-defined distance threshold $\tau$ are assigned a density value of 1, while those that do not meet the threshold are assigned a density value of 0. $\tau$ is set to 0.01 in practice. The density threshold for Marching-Cube is set to 1e-4. 

Based on the visual comparison, the overall mesh quality can be ranked as follows: NeuS $>$ Poisson Reconstruction $>$ Marching-Cube.

\subsection*{C.2 Training Details of Gaussian-Binding Stage}
To ensure an accurate representation of each triangle, we bind  $N$ Gaussians to it. Prior to training, we initialize the positions of the Gaussians on the attached triangle. The $N$ initialized position is calculated using a barycentric coordinate, with a predefined barycentric coordinate set of $[1/2, 1/4, 1/4]$, $[1/4, 1/2, 1/4]$, $[1/4, 1/4, 1/2]$. For the hyper-parameter $\beta$ mentioned in main paper equation \textcolor{red}{\textbf{{(8)}}}, we set $\beta=10$ in most cases, $\beta=100$ in \textit{Materials}.

In the Gaussian-Binding stage, we don't perform adaptive control because we find it doesn't influence the final performance. We also train 300K iterations in this stage with L1 loss, SSIM loss, and mask entropy loss. The overall loss in this can be summarized as follows:
\begin{equation}
\begin{split}
    \mathcal{L}_{stage2} = \lambda_1\mathcal{L}_1 + \lambda_2\mathcal{L}_{SSIM} + \lambda_3\mathcal{L}_{mask},
\end{split} 
\end{equation}
where $\lambda_1=1, \lambda_2=0.2 ,\lambda_3=0.1$.

%% file: main.bbl
\begin{thebibliography}{60}
\providecommand{\natexlab}[1]{#1}
\providecommand{\url}[1]{\texttt{#1}}
\expandafter\ifx\csname urlstyle\endcsname\relax
  \providecommand{\doi}[1]{doi: #1}\else
  \providecommand{\doi}{doi: \begingroup \urlstyle{rm}\Url}\fi

\bibitem[Abdal et~al.(2024)Abdal, Yifan, Shi, Xu, Po, Kuang, Chen, Yeung, and Wetzstein]{abdal2024gaussianshellmap}
Rameen Abdal, Wang Yifan, Zifan Shi, Yinghao Xu, Ryan Po, Zhengfei Kuang, Qifeng Chen, Dit-Yan Yeung, and Gordon Wetzstein.
\newblock Gaussian shell maps for efficient 3d human generation.
\newblock In \emph{Proceedings of the IEEE/CVF Conference on Computer Vision and Pattern Recognition}, pages 9441--9451, 2024.

\bibitem[Bao et~al.(2023)Bao, Zhang, Yang, Fan, Yang, Bao, Zhang, and Cui]{bao2023sine}
Chong Bao, Yinda Zhang, Bangbang Yang, Tianxing Fan, Zesong Yang, Hujun Bao, Guofeng Zhang, and Zhaopeng Cui.
\newblock Sine: Semantic-driven image-based nerf editing with prior-guided editing field.
\newblock In \emph{Proceedings of the IEEE/CVF Conference on Computer Vision and Pattern Recognition}, pages 20919--20929, 2023.

\bibitem[Chen et~al.(2023{\natexlab{a}})Chen, Wang, Chen, Zhang, Li, Guo, Wang, and Wang]{chen2023uv}
Yue Chen, Xuan Wang, Xingyu Chen, Qi Zhang, Xiaoyu Li, Yu Guo, Jue Wang, and Fei Wang.
\newblock Uv volumes for real-time rendering of editable free-view human performance.
\newblock In \emph{Proceedings of the IEEE/CVF Conference on Computer Vision and Pattern Recognition}, pages 16621--16631, 2023{\natexlab{a}}.

\bibitem[Chen et~al.(2023{\natexlab{b}})Chen, Funkhouser, Hedman, and Tagliasacchi]{chen2023mobilenerf}
Zhiqin Chen, Thomas Funkhouser, Peter Hedman, and Andrea Tagliasacchi.
\newblock Mobilenerf: Exploiting the polygon rasterization pipeline for efficient neural field rendering on mobile architectures.
\newblock In \emph{Proceedings of the IEEE/CVF Conference on Computer Vision and Pattern Recognition}, pages 16569--16578, 2023{\natexlab{b}}.

\bibitem[Feng et~al.(2024)Feng, Feng, Shang, Jiang, Yu, Zong, Shao, Wu, Zhou, Jiang, et~al.]{feng2024gaussian}
Yutao Feng, Xiang Feng, Yintong Shang, Ying Jiang, Chang Yu, Zeshun Zong, Tianjia Shao, Hongzhi Wu, Kun Zhou, Chenfanfu Jiang, et~al.
\newblock Gaussian splashing: Dynamic fluid synthesis with gaussian splatting.
\newblock \emph{arXiv preprint arXiv:2401.15318}, 2024.

\bibitem[Gao et~al.(2023)Gao, Gu, Lin, Zhu, Cao, Zhang, and Yao]{gao2023relightable}
Jian Gao, Chun Gu, Youtian Lin, Hao Zhu, Xun Cao, Li Zhang, and Yao Yao.
\newblock Relightable 3d gaussian: Real-time point cloud relighting with brdf decomposition and ray tracing.
\newblock \emph{arXiv preprint arXiv:2311.16043}, 2023.

\bibitem[Gao et~al.(2024{\natexlab{a}})Gao, Yang, Zhang, Sun, Yuan, Fu, and Lai]{gao2024mesh}
Lin Gao, Jie Yang, Bo-Tao Zhang, Jia-Mu Sun, Yu-Jie Yuan, Hongbo Fu, and Yu-Kun Lai.
\newblock Mesh-based gaussian splatting for real-time large-scale deformation.
\newblock \emph{arXiv preprint arXiv:2402.04796}, 2024{\natexlab{a}}.

\bibitem[Gao et~al.(2022)Gao, Yang, Kim, Peng, Liu, and Tong]{gao2022mps}
Xiangjun Gao, Jiaolong Yang, Jongyoo Kim, Sida Peng, Zicheng Liu, and Xin Tong.
\newblock Mps-nerf: Generalizable 3d human rendering from multiview images.
\newblock \emph{IEEE Transactions on Pattern Analysis and Machine Intelligence}, 2022.

\bibitem[Gao et~al.(2024{\natexlab{b}})Gao, Li, Zhang, Zhang, Cao, Shan, and Quan]{gao2024contex}
Xiangjun Gao, Xiaoyu Li, Chaopeng Zhang, Qi Zhang, Yanpei Cao, Ying Shan, and Long Quan.
\newblock Contex-human: Free-view rendering of human from a single image with texture-consistent synthesis.
\newblock In \emph{Proceedings of the IEEE/CVF Conference on Computer Vision and Pattern Recognition}, pages 10084--10094, 2024{\natexlab{b}}.

\bibitem[Gu{\'e}don and Lepetit(2023)]{guedon2023sugar}
Antoine Gu{\'e}don and Vincent Lepetit.
\newblock Sugar: Surface-aligned gaussian splatting for efficient 3d mesh reconstruction and high-quality mesh rendering.
\newblock \emph{arXiv preprint arXiv:2311.12775}, 2023.

\bibitem[Gu{\'e}don and Lepetit(2024)]{guedon2024gaussian}
Antoine Gu{\'e}don and Vincent Lepetit.
\newblock Gaussian frosting: Editable complex radiance fields with real-time rendering.
\newblock \emph{arXiv preprint arXiv:2403.14554}, 2024.

\bibitem[He et~al.(2024)He, Li, Kang, Ye, Zhang, Chen, Gao, Zhang, Wu, and Zhuang]{he2024magicman}
Xu He, Xiaoyu Li, Di Kang, Jiangnan Ye, Chaopeng Zhang, Liyang Chen, Xiangjun Gao, Han Zhang, Zhiyong Wu, and Haolin Zhuang.
\newblock Magicman: Generative novel view synthesis of humans with 3d-aware diffusion and iterative refinement.
\newblock \emph{arXiv preprint arXiv:2408.14211}, 2024.

\bibitem[Hu et~al.(2023)Hu, Zhang, Zhang, Zhou, Liu, Zhang, and Nie]{hu2023gaussianavatar}
Liangxiao Hu, Hongwen Zhang, Yuxiang Zhang, Boyao Zhou, Boning Liu, Shengping Zhang, and Liqiang Nie.
\newblock Gaussianavatar: Towards realistic human avatar modeling from a single video via animatable 3d gaussians.
\newblock \emph{arXiv preprint arXiv:2312.02134}, 2023.

\bibitem[Huang et~al.(2023)Huang, Sun, Yang, Lyu, Cao, and Qi]{huang2023sc}
Yi-Hua Huang, Yang-Tian Sun, Ziyi Yang, Xiaoyang Lyu, Yan-Pei Cao, and Xiaojuan Qi.
\newblock Sc-gs: Sparse-controlled gaussian splatting for editable dynamic scenes.
\newblock \emph{arXiv preprint arXiv:2312.14937}, 2023.

\bibitem[Jambon et~al.(2023)Jambon, Kerbl, Kopanas, Diolatzis, Drettakis, and Leimk{\"u}hler]{jambon2023nerfshop}
Cl{\'e}ment Jambon, Bernhard Kerbl, Georgios Kopanas, Stavros Diolatzis, George Drettakis, and Thomas Leimk{\"u}hler.
\newblock Nerfshop: Interactive editing of neural radiance fields.
\newblock \emph{Proceedings of the ACM on Computer Graphics and Interactive Techniques}, 6\penalty0 (1), 2023.

\bibitem[Jensen et~al.(2014)Jensen, Dahl, Vogiatzis, Tola, and Aan{\ae}s]{jensen2014dtu}
Rasmus Jensen, Anders Dahl, George Vogiatzis, Engin Tola, and Henrik Aan{\ae}s.
\newblock Large scale multi-view stereopsis evaluation.
\newblock In \emph{Proceedings of the IEEE conference on computer vision and pattern recognition}, pages 406--413, 2014.

\bibitem[Jiang et~al.(2024)Jiang, Yu, Xie, Li, Feng, Wang, Li, Lau, Gao, Yang, et~al.]{jiang2024vr}
Ying Jiang, Chang Yu, Tianyi Xie, Xuan Li, Yutao Feng, Huamin Wang, Minchen Li, Henry Lau, Feng Gao, Yin Yang, et~al.
\newblock Vr-gs: A physical dynamics-aware interactive gaussian splatting system in virtual reality.
\newblock \emph{arXiv preprint arXiv:2401.16663}, 2024.

\bibitem[Kazhdan and Hoppe(2013)]{kazhdan2013screened}
Michael Kazhdan and Hugues Hoppe.
\newblock Screened poisson surface reconstruction.
\newblock \emph{ACM Transactions on Graphics (ToG)}, 32\penalty0 (3):\penalty0 1--13, 2013.

\bibitem[Kerbl et~al.(2023)Kerbl, Kopanas, Leimk{\"u}hler, and Drettakis]{kerbl20233d}
Bernhard Kerbl, Georgios Kopanas, Thomas Leimk{\"u}hler, and George Drettakis.
\newblock 3d gaussian splatting for real-time radiance field rendering.
\newblock \emph{ACM Transactions on Graphics}, 42\penalty0 (4), 2023.

\bibitem[Kirschstein et~al.(2023)Kirschstein, Giebenhain, and Nie{\ss}ner]{kirschstein2023diffusionavatars}
Tobias Kirschstein, Simon Giebenhain, and Matthias Nie{\ss}ner.
\newblock Diffusionavatars: Deferred diffusion for high-fidelity 3d head avatars.
\newblock \emph{arXiv preprint arXiv:2311.18635}, 2023.

\bibitem[Knapitsch et~al.(2017)Knapitsch, Park, Zhou, and Koltun]{Knapitsch2017_tnt}
Arno Knapitsch, Jaesik Park, Qian-Yi Zhou, and Vladlen Koltun.
\newblock Tanks and temples: Benchmarking large-scale scene reconstruction.
\newblock \emph{ACM Transactions on Graphics}, 36\penalty0 (4), 2017.

\bibitem[Liang et~al.(2023)Liang, Zhang, Feng, Shan, and Jia]{liang2023gs}
Zhihao Liang, Qi Zhang, Ying Feng, Ying Shan, and Kui Jia.
\newblock Gs-ir: 3d gaussian splatting for inverse rendering.
\newblock \emph{arXiv preprint arXiv:2311.16473}, 2023.

\bibitem[Lin et~al.(2023)Lin, Dai, Zhu, and Yao]{lin2023gaussian}
Youtian Lin, Zuozhuo Dai, Siyu Zhu, and Yao Yao.
\newblock Gaussian-flow: 4d reconstruction with dynamic 3d gaussian particle.
\newblock \emph{arXiv preprint arXiv:2312.03431}, 2023.

\bibitem[Liu et~al.(2023)Liu, Xiang, Zhao, Zhang, Yu, and Zheng]{liu2023neural}
Ruiyang Liu, Jinxu Xiang, Bowen Zhao, Ran Zhang, Jingyi Yu, and Changxi Zheng.
\newblock Neural impostor: Editing neural radiance fields with explicit shape manipulation.
\newblock In \emph{Computer Graphics Forum}, page e14981. Wiley Online Library, 2023.

\bibitem[Liu et~al.(2021)Liu, Zhang, Zhang, Zhang, Zhu, and Russell]{liu2021editing}
Steven Liu, Xiuming Zhang, Zhoutong Zhang, Richard Zhang, Jun-Yan Zhu, and Bryan Russell.
\newblock Editing conditional radiance fields.
\newblock In \emph{Proceedings of the IEEE/CVF international conference on computer vision}, pages 5773--5783, 2021.

\bibitem[Ma et~al.(2022)Ma, Li, Liao, Wang, Zhang, Wang, and Sander]{ma2022neural}
Li Ma, Xiaoyu Li, Jing Liao, Xuan Wang, Qi Zhang, Jue Wang, and Pedro~V Sander.
\newblock Neural parameterization for dynamic human head editing.
\newblock \emph{ACM Transactions on Graphics (TOG)}, 41\penalty0 (6):\penalty0 1--15, 2022.

\bibitem[Mildenhall et~al.(2021)Mildenhall, Srinivasan, Tancik, Barron, Ramamoorthi, and Ng]{mildenhall2021nerf}
Ben Mildenhall, Pratul~P Srinivasan, Matthew Tancik, Jonathan~T Barron, Ravi Ramamoorthi, and Ren Ng.
\newblock Nerf: Representing scenes as neural radiance fields for view synthesis.
\newblock \emph{Communications of the ACM}, 65\penalty0 (1):\penalty0 99--106, 2021.

\bibitem[Qian et~al.(2023)Qian, Kirschstein, Schoneveld, Davoli, Giebenhain, and Nie{\ss}ner]{qian2023gaussianavatars}
Shenhan Qian, Tobias Kirschstein, Liam Schoneveld, Davide Davoli, Simon Giebenhain, and Matthias Nie{\ss}ner.
\newblock Gaussianavatars: Photorealistic head avatars with rigged 3d gaussians.
\newblock \emph{arXiv preprint arXiv:2312.02069}, 2023.

\bibitem[Radford et~al.(2021)Radford, Kim, Hallacy, Ramesh, Goh, Agarwal, Sastry, Askell, Mishkin, Clark, et~al.]{radford2021learning}
Alec Radford, Jong~Wook Kim, Chris Hallacy, Aditya Ramesh, Gabriel Goh, Sandhini Agarwal, Girish Sastry, Amanda Askell, Pamela Mishkin, Jack Clark, et~al.
\newblock Learning transferable visual models from natural language supervision.
\newblock In \emph{International conference on machine learning}, pages 8748--8763. PMLR, 2021.

\bibitem[Rakotosaona et~al.(2023)Rakotosaona, Manhardt, Arroyo, Niemeyer, Kundu, and Tombari]{rakotosaona2023nerfmeshing}
Marie-Julie Rakotosaona, Fabian Manhardt, Diego~Martin Arroyo, Michael Niemeyer, Abhijit Kundu, and Federico Tombari.
\newblock Nerfmeshing: Distilling neural radiance fields into geometrically-accurate 3d meshes.
\newblock \emph{arXiv preprint arXiv:2303.09431}, 2023.

\bibitem[Sun et~al.(2022)Sun, Wang, Zhang, Li, Zhang, Liu, and Wang]{sun2022fenerf}
Jingxiang Sun, Xuan Wang, Yong Zhang, Xiaoyu Li, Qi Zhang, Yebin Liu, and Jue Wang.
\newblock Fenerf: Face editing in neural radiance fields.
\newblock In \emph{Proceedings of the IEEE/CVF Conference on Computer Vision and Pattern Recognition}, pages 7672--7682, 2022.

\bibitem[Tang et~al.(2023{\natexlab{a}})Tang, Ren, Zhou, Liu, and Zeng]{tang2023dreamgaussian}
Jiaxiang Tang, Jiawei Ren, Hang Zhou, Ziwei Liu, and Gang Zeng.
\newblock Dreamgaussian: Generative gaussian splatting for efficient 3d content creation.
\newblock \emph{arXiv preprint arXiv:2309.16653}, 2023{\natexlab{a}}.

\bibitem[Tang et~al.(2023{\natexlab{b}})Tang, Zhou, Chen, Hu, Ding, Wang, and Zeng]{tang2023delicate}
Jiaxiang Tang, Hang Zhou, Xiaokang Chen, Tianshu Hu, Errui Ding, Jingdong Wang, and Gang Zeng.
\newblock Delicate textured mesh recovery from nerf via adaptive surface refinement.
\newblock \emph{arXiv preprint arXiv:2303.02091}, 2023{\natexlab{b}}.

\bibitem[Waczy{\'n}ska et~al.(2024)Waczy{\'n}ska, Borycki, Tadeja, Tabor, and Spurek]{waczynska2024games}
Joanna Waczy{\'n}ska, Piotr Borycki, S{\l}awomir Tadeja, Jacek Tabor, and Przemys{\l}aw Spurek.
\newblock Games: Mesh-based adapting and modification of gaussian splatting.
\newblock \emph{arXiv preprint arXiv:2402.01459}, 2024.

\bibitem[Wang et~al.(2022)Wang, Chai, He, Chen, and Liao]{wang2022clip}
Can Wang, Menglei Chai, Mingming He, Dongdong Chen, and Jing Liao.
\newblock Clip-nerf: Text-and-image driven manipulation of neural radiance fields.
\newblock In \emph{Proceedings of the IEEE/CVF Conference on Computer Vision and Pattern Recognition}, pages 3835--3844, 2022.

\bibitem[Wang et~al.(2023{\natexlab{a}})Wang, He, Chai, Chen, and Liao]{wang2023mesh}
Can Wang, Mingming He, Menglei Chai, Dongdong Chen, and Jing Liao.
\newblock Mesh-guided neural implicit field editing.
\newblock \emph{arXiv preprint arXiv:2312.02157}, 2023{\natexlab{a}}.

\bibitem[Wang et~al.(2021)Wang, Liu, Liu, Theobalt, Komura, and Wang]{wang2021neus}
Peng Wang, Lingjie Liu, Yuan Liu, Christian Theobalt, Taku Komura, and Wenping Wang.
\newblock Neus: Learning neural implicit surfaces by volume rendering for multi-view reconstruction.
\newblock \emph{arXiv preprint arXiv:2106.10689}, 2021.

\bibitem[Wang et~al.(2023{\natexlab{b}})Wang, Zhu, Ye, Huo, Ran, Zhong, and Chen]{wang2023seal3d}
Xiangyu Wang, Jingsen Zhu, Qi Ye, Yuchi Huo, Yunlong Ran, Zhihua Zhong, and Jiming Chen.
\newblock Seal-3d: Interactive pixel-level editing for neural radiance fields, 2023{\natexlab{b}}.

\bibitem[Wu et~al.(2023{\natexlab{a}})Wu, Yi, Fang, Xie, Zhang, Wei, Liu, Tian, and Wang]{wu20234d}
Guanjun Wu, Taoran Yi, Jiemin Fang, Lingxi Xie, Xiaopeng Zhang, Wei Wei, Wenyu Liu, Qi Tian, and Xinggang Wang.
\newblock 4d gaussian splatting for real-time dynamic scene rendering.
\newblock \emph{arXiv preprint arXiv:2310.08528}, 2023{\natexlab{a}}.

\bibitem[Wu et~al.(2023{\natexlab{b}})Wu, Zhu, Huang, Zhuang, Lu, and Cao]{wu2023high}
Menghua Wu, Hao Zhu, Linjia Huang, Yiyu Zhuang, Yuanxun Lu, and Xun Cao.
\newblock High-fidelity 3d face generation from natural language descriptions.
\newblock In \emph{Proceedings of the IEEE/CVF Conference on Computer Vision and Pattern Recognition}, pages 4521--4530, 2023{\natexlab{b}}.

\bibitem[Xiang et~al.(2021)Xiang, Xu, Hasan, Hold-Geoffroy, Sunkavalli, and Su]{xiang2021neutex}
Fanbo Xiang, Zexiang Xu, Milos Hasan, Yannick Hold-Geoffroy, Kalyan Sunkavalli, and Hao Su.
\newblock Neutex: Neural texture mapping for volumetric neural rendering.
\newblock In \emph{Proceedings of the IEEE/CVF Conference on Computer Vision and Pattern Recognition}, pages 7119--7128, 2021.

\bibitem[Xie et~al.(2023)Xie, Zong, Qiu, Li, Feng, Yang, and Jiang]{xie2023physgaussian}
Tianyi Xie, Zeshun Zong, Yuxin Qiu, Xuan Li, Yutao Feng, Yin Yang, and Chenfanfu Jiang.
\newblock Physgaussian: Physics-integrated 3d gaussians for generative dynamics.
\newblock \emph{arXiv preprint arXiv:2311.12198}, 2023.

\bibitem[Xu et~al.(2022)Xu, Xu, Philip, Bi, Shu, Sunkavalli, and Neumann]{xu2022point}
Qiangeng Xu, Zexiang Xu, Julien Philip, Sai Bi, Zhixin Shu, Kalyan Sunkavalli, and Ulrich Neumann.
\newblock Point-nerf: Point-based neural radiance fields.
\newblock In \emph{Proceedings of the IEEE/CVF Conference on Computer Vision and Pattern Recognition}, pages 5438--5448, 2022.

\bibitem[Xu et~al.(2023)Xu, Chen, Li, Zhang, Wang, Zheng, and Liu]{xu2023gaussian}
Yuelang Xu, Benwang Chen, Zhe Li, Hongwen Zhang, Lizhen Wang, Zerong Zheng, and Yebin Liu.
\newblock Gaussian head avatar: Ultra high-fidelity head avatar via dynamic gaussians.
\newblock \emph{arXiv preprint arXiv:2312.03029}, 2023.

\bibitem[Yang et~al.(2022)Yang, Bao, Zeng, Bao, Zhang, Cui, and Zhang]{yang2022neumesh}
Bangbang Yang, Chong Bao, Junyi Zeng, Hujun Bao, Yinda Zhang, Zhaopeng Cui, and Guofeng Zhang.
\newblock Neumesh: Learning disentangled neural mesh-based implicit field for geometry and texture editing.
\newblock In \emph{European Conference on Computer Vision}, pages 597--614. Springer, 2022.

\bibitem[Yang et~al.(2023)Yang, Gao, Zhou, Jiao, Zhang, and Jin]{yang2023deformable}
Ziyi Yang, Xinyu Gao, Wen Zhou, Shaohui Jiao, Yuqing Zhang, and Xiaogang Jin.
\newblock Deformable 3d gaussians for high-fidelity monocular dynamic scene reconstruction.
\newblock \emph{arXiv preprint arXiv:2309.13101}, 2023.

\bibitem[Yao et~al.(2022)Yao, Zhang, Liu, Qu, Fang, McKinnon, Tsin, and Quan]{yao2022neilf}
Yao Yao, Jingyang Zhang, Jingbo Liu, Yihang Qu, Tian Fang, David McKinnon, Yanghai Tsin, and Long Quan.
\newblock Neilf: Neural incident light field for physically-based material estimation.
\newblock In \emph{European Conference on Computer Vision}, pages 700--716. Springer, 2022.

\bibitem[Yariv et~al.(2023)Yariv, Hedman, Reiser, Verbin, Srinivasan, Szeliski, Barron, and Mildenhall]{yariv2023bakedsdf}
Lior Yariv, Peter Hedman, Christian Reiser, Dor Verbin, Pratul~P Srinivasan, Richard Szeliski, Jonathan~T Barron, and Ben Mildenhall.
\newblock Bakedsdf: Meshing neural sdfs for real-time view synthesis.
\newblock \emph{arXiv preprint arXiv:2302.14859}, 2023.

\bibitem[Yu et~al.(2024)Yu, Yuan, Cao, Gao, Li, Hu, Quan, Shan, and Tian]{yu2024hifi}
Wangbo Yu, Li Yuan, Yan-Pei Cao, Xiangjun Gao, Xiaoyu Li, Wenbo Hu, Long Quan, Ying Shan, and Yonghong Tian.
\newblock Hifi-123: Towards high-fidelity one image to 3d content generation.
\newblock In \emph{European Conference on Computer Vision}, pages 258--274. Springer, 2024.

\bibitem[Yuan et~al.(2023)Yuan, Li, Huang, De~Mello, Nagano, Kautz, and Iqbal]{yuan2023gavatar}
Ye Yuan, Xueting Li, Yangyi Huang, Shalini De~Mello, Koki Nagano, Jan Kautz, and Umar Iqbal.
\newblock Gavatar: Animatable 3d gaussian avatars with implicit mesh learning.
\newblock \emph{arXiv preprint arXiv:2312.11461}, 2023.

\bibitem[Yuan et~al.(2022)Yuan, Sun, Lai, Ma, Jia, and Gao]{yuan2022nerf}
Yu-Jie Yuan, Yang-Tian Sun, Yu-Kun Lai, Yuewen Ma, Rongfei Jia, and Lin Gao.
\newblock Nerf-editing: geometry editing of neural radiance fields.
\newblock In \emph{Proceedings of the IEEE/CVF Conference on Computer Vision and Pattern Recognition}, pages 18353--18364, 2022.

\bibitem[Zhan et~al.(2023)Zhan, Liu, Kortylewski, and Theobalt]{zhan2023general}
Fangneng Zhan, Lingjie Liu, Adam Kortylewski, and Christian Theobalt.
\newblock General neural gauge fields.
\newblock \emph{arXiv preprint arXiv:2305.03462}, 2023.

\bibitem[Zhang et~al.(2022)Zhang, Li, Wan, Wang, and Liao]{zhang2022fdnerf}
Jingbo Zhang, Xiaoyu Li, Ziyu Wan, Can Wang, and Jing Liao.
\newblock Fdnerf: Few-shot dynamic neural radiance fields for face reconstruction and expression editing.
\newblock In \emph{SIGGRAPH Asia 2022 Conference Papers}, pages 1--9, 2022.

\bibitem[Zheng et~al.(2022)Zheng, Huang, Yu, Zhang, Guo, and Liu]{zheng2022structured}
Zerong Zheng, Han Huang, Tao Yu, Hongwen Zhang, Yandong Guo, and Yebin Liu.
\newblock Structured local radiance fields for human avatar modeling.
\newblock In \emph{Proceedings of the IEEE/CVF Conference on Computer Vision and Pattern Recognition}, pages 15893--15903, 2022.

\bibitem[Zhou et~al.(2023{\natexlab{a}})Zhou, Hong, Xie, Yang, Li, and Zhang]{zhou2023serf}
Kaichen Zhou, Lanqing Hong, Enze Xie, Yongxin Yang, Zhenguo Li, and Wei Zhang.
\newblock Serf: Fine-grained interactive 3d segmentation and editing with radiance fields.
\newblock \emph{arXiv preprint arXiv:2312.15856}, 2023{\natexlab{a}}.

\bibitem[Zhou et~al.(2023{\natexlab{b}})Zhou, Lin, Shan, Wang, Sun, and Yang]{zhou2023drivinggaussian}
Xiaoyu Zhou, Zhiwei Lin, Xiaojun Shan, Yongtao Wang, Deqing Sun, and Ming-Hsuan Yang.
\newblock Drivinggaussian: Composite gaussian splatting for surrounding dynamic autonomous driving scenes.
\newblock \emph{arXiv preprint arXiv:2312.07920}, 2023{\natexlab{b}}.

\bibitem[Zhuang et~al.(2022)Zhuang, Zhu, Sun, and Cao]{zhuang2022mofanerf}
Yiyu Zhuang, Hao Zhu, Xusen Sun, and Xun Cao.
\newblock Mofanerf: Morphable facial neural radiance field.
\newblock In \emph{European conference on computer vision}, pages 268--285. Springer, 2022.

\bibitem[Zhuang et~al.(2024)Zhuang, He, Zhang, Wang, Zhu, Yao, Zhu, Cao, and Zhu]{zhuang2024towards}
Yiyu Zhuang, Yuxiao He, Jiawei Zhang, Yanwen Wang, Jiahe Zhu, Yao Yao, Siyu Zhu, Xun Cao, and Hao Zhu.
\newblock Towards native generative model for 3d head avatar.
\newblock \emph{arXiv preprint arXiv:2410.01226}, 2024.

\bibitem[Zielonka et~al.(2023)Zielonka, Bagautdinov, Saito, Zollh{\"o}fer, Thies, and Romero]{zielonka2023drivable}
Wojciech Zielonka, Timur Bagautdinov, Shunsuke Saito, Michael Zollh{\"o}fer, Justus Thies, and Javier Romero.
\newblock Drivable 3d gaussian avatars.
\newblock \emph{arXiv preprint arXiv:2311.08581}, 2023.

\bibitem[Zwicker et~al.(2002)Zwicker, Pfister, Van~Baar, and Gross]{zwicker2002ewa}
Matthias Zwicker, Hanspeter Pfister, Jeroen Van~Baar, and Markus Gross.
\newblock Ewa splatting.
\newblock \emph{IEEE Transactions on Visualization and Computer Graphics}, 8\penalty0 (3):\penalty0 223--238, 2002.

\end{thebibliography}
